\documentclass[12pt,epsfig]{article}
\usepackage{epsf}
\usepackage{graphicx}
\usepackage{eepic}
\usepackage{color}
\usepackage{latexsym,epsfig}
\newcommand{\Frac}[2]{\frac{\displaystyle #1}{\displaystyle #2}}
\newcommand{\Int}{\displaystyle{\int}}

\begin{document}

\begin{center}
{\Large\bf Studies on $\pi^+\pi^-$ phase motion in the
$\Psi'\rightarrow J/\Psi \pi^+ \pi^-$ process }
\\[10mm]
{\sc Z.~X.~Zhang$^{1,2}$, J.~J.~Sanz-Cillero$^3$, X.~Y.~Shen$^2$
N.~Wu$^2$, L.~Y.~Xiao$^4$,
 and H.~Q.~Zheng$^1$}
\\[2mm]
{\it 1) Department of Physics, Peking University, Beijing 100871,
	P.~R.~China}
\\[2mm]
{\it 2) Institute of High Energy Physics, Chinese Academy of
	Science, Beijing 100039, P.~R.~China}
\\ [2mm]{ \it 3) IFAE, Universitat Autonoma de Barcelona, 08193 Bellaterra, 
	Barcelona, Spain}
\\ [2mm]{ \it 4) Olivet Institute of Technology, Olivet University, 
	San Francisco, CA 94103, USA }
\\[5mm]
\today

\end{center}

\vskip 1cm
\begin{abstract}
We propose a measurement on the elastic $\pi\pi$ scattering phase
shift difference $\delta^0_0-\delta^2_0$ through $\Psi'\rightarrow
J/\Psi\, \pi^+ \pi^-$ process in future high statistics BES-III
experiment. The decay amplitude is constructed with seven Lorentz
invariant form-factors and is compared with their  theoretical
estimation.
 It is found that
the phase shift difference can be obtained, based on a Monte Carlo
study and it is expected the phase shift in the energy region
between 350 MeV to 550 MeV can be  measured at future BES-III.
\end{abstract}

 \section{Introduction}
In recent years, the operation of a number of high precision, high
statistics experimental machines, varying from fixed target
experiments to $B$ factories, opens a new era for precision hadronic
experiments. Based on that, both experimental and theoretical
studies on low energy $\pi\pi$ and $\pi K$ system produced in
production processes have also received revived interests. The
importance of these studies follows from the fact that when final
state theorem applies, one can extract  low energy elastic
scattering phase shifts in the related channels through a partial
wave analysis. The information on $\pi\pi$, $\pi K$ phase shifts
then provides a crucial ingredient in understanding the dynamics of
goldstone bosons and the spontaneous breaking of chiral symmetry.

The experimental and theoretical activities in the last few years
mainly focused on semi-leptonic and hadronic $D$ decays~(see for
example, Refs.~\cite{Pennington} -- \cite{caprini}). It is known
that in $D$ semi-leptonic decays $p$--wave  dominates, and the more
interesting $s$--wave component is small. In this paper we
re-investigate the $\pi\pi$ final state interactions in the
$\Psi'\rightarrow J/\Psi \pi^+ \pi^-$ process.  Here $s$--wave
dominates and the next contribution comes from the tiny $d$--wave.
The existence of the latter is however crucial for exploring the
$s$--wave phase motion through interference effect. The decay
product under concern is a three body final state, the $J/\Psi$
particle is however irrelevant to any final state interactions here.
Using color transparency argument it is understood that the effect
from rescattering between the $J/\Psi$ and one of the pions is
negligible. Another important fact is that, in the kinematic region
under concern, between the initial $\Psi'$ and the final
$J/\Psi\pi\pi$ there is no other on-shell intermediate hadronic
state available (or are double OZI suppressed and hence negligible).
Hence the final state theorem is applicable to the $\pi\pi$ system
in $\Psi'\rightarrow J/\Psi \pi^+ \pi^-$ process.

The $\Psi'\rightarrow J/\Psi \pi^+ \pi^-$  process has been the
subject of a number of previous publications~(\cite{Jpsipipi}--
\cite{Jpsipipith}).  In Ref.~\cite{Cahn:1975ts}, the author proposed
a method to extract $\pi\pi$ phase shift from $\Psi'\rightarrow
J/\Psi \pi^+ \pi^-$, similar to Pais-Treiman~\cite{Pais} method for
obtaining $\pi\pi$ phase shifts from $K_{l4}$ decays, but only
considered three partial wave amplitudes for reducing the difficulty
of the analysis. A similar method~\cite{PKo} was also proposed in
$\Upsilon(3S)\to\Upsilon(1S)\pi\pi$ process but only the lowest
order in the pion momentum expansion was considered. In this work we
are able to provide a more general parametrization to the decay
amplitude comparing with what is given in most of previous papers.
Our parametrization will be discussed in section 2. Furthermore we
will also provide a Monte Carlo study in section 3 to test the
stability and reliability to use our parametrization to extract the
phase shift data.

\section{General structure of the $\Psi'\rightarrow J/\Psi\, \pi^+ \pi^-$ decay amplitude}
\subsection{The Lorentz invariant form-factors}
There are 3 independent momenta $p_{\pi^+}, p_{\pi^-},
p_{J/\Psi}=p_3$, which can be re-expressed in 3 variables
$q=p_{\pi^+}+p_{\pi^-}, p=p_{\pi^+}-p_{\pi^-},$ and
 $p_3$. The three independent momenta can form 2 independent Lorentz invariant products,  chosen as $q^2$ and $p\cdot
 p_3$ here. Then,
\begin{eqnarray}
&&q^2=s,\\
&&p^2=-s \rho^2=4m_\pi^2-s ,\\
&&p_3^2=M_\Psi^2,\\
&&q\cdot p=0,\\
&&q\cdot p_3=\frac{1}{2}(M_{\Psi'}^2-M_\Psi^2-s),
\end{eqnarray}
with the kinematics factor of dipion system
$\rho=\sqrt{1-{4m_\pi^2\over s}}$,  the energy and the momenta of
$J/\Psi$ in the lab frame (the $\Psi'$ rest frame), which
$E_3=(M_{\Psi'}^2+M_\Psi^2-s)/(2M_{\Psi'})$ and $
|\vec{p}_3|=\sqrt{E_3^2-M_\Psi^2}$ are functions of $q^2$.
Moreover, $p\cdot p_3$ can be expressed in the $\Psi'$ rest frame
with the variables in the dipion rest frame,
\begin{equation}
p\cdot p_3=2\gamma |\vec{p}_{\pi\pi}^{\,\,*}|\cos\theta_\pi^*(\beta
E_3+|\vec{p}_3|),
\end{equation}
where $|\vec{p}_{\pi\pi}^{\,\,*}|=\rho \sqrt{s}/2$ is the three momenta
of $\pi$ in the dipion rest frame and $\theta_\pi^*$ is the angle
between the direction of $\pi^+$ and direction opposite to the
final $J/\Psi$ in the dipion rest frame, see Fig.~1 for
illustration. $\beta=\sqrt{1-\frac{1}{\gamma^2}}$ is the boost
factor from the dipion rest frame to the lab frame with
$\gamma=(M_{\Psi'}^2-M_\Psi^2+s)/(2M_{\Psi'}\sqrt{s})$.

\begin{figure}
\begin{center}\epsfxsize=6cm\epsfbox{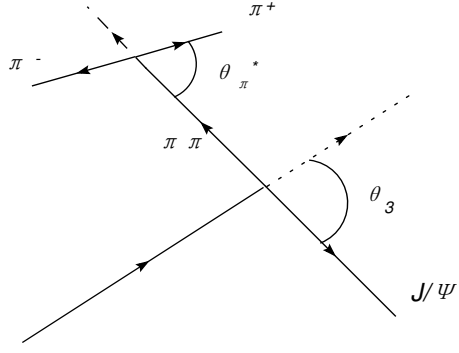}
\caption{$\theta_3$ is the angle
between the beam and $J/\Psi$ in the $\Psi'$ rest frame.
$\theta_\pi^*$ is  the angle between the $J/\Psi$ direction and the
$\pi^+$ in the di-pion rest frame,  $\phi$ is the azimuthal angle
between the  beam  $J/\Psi$ plane and $\pi^+\pi^-$ plane in the
$\Psi'$ rest frame (not drawn in the figure). }
\end{center}\end{figure}

Denoting the polarization vector of $\Psi'$ and $J/\Psi$ by
$\epsilon'$ and $\epsilon$ respectively, we can form five
invariants bilinear in
$\epsilon'\epsilon^*$:
\[(\epsilon\cdot\epsilon'),\,\,\,(\epsilon\cdot q)(
\epsilon' \cdot q),\,\,\,(\epsilon\cdot p)( \epsilon'\cdot
p),\,\,\, (\epsilon \cdot q)( \epsilon' \cdot p),\,\,\,(\epsilon
\cdot p)( \epsilon' \cdot q).\] Hence, the amplitude have the
independent structure:
\begin{equation}
T=(\epsilon\cdot\epsilon')F_0+(\epsilon\cdot q)( \epsilon' \cdot
q)F_1+(\epsilon\cdot p)( \epsilon'\cdot p)F_2+(\epsilon \cdot q)(
\epsilon' \cdot p)F_3+(\epsilon \cdot p)( \epsilon' \cdot q)F_4.
\end{equation}
$F_i$ are the functions of $s$ and $p\cdot p_3$ and these allow us
to do partial wave decomposition according to $\theta_\pi^*$.

\subsection{Partial wave decomposition} From Ref.~\cite{Chung:93},
we obtain the basis of tensors
\begin{eqnarray}
&&\tilde{t}^{(0)}=1,\\
&&\tilde{t}^{(1)}=p^\mu,\\
&&\tilde{t}^{(2)}=p^\mu p^\nu -\frac{1}{3}p^2 \tilde{g}^{\mu\nu},
\,\,\,\,\hbox{with}\,\,\,\tilde{g}^{\mu\nu}=g^{\mu\nu}-\frac{q^\mu
q^\nu}{q^2},
\end{eqnarray}
where every tensor $\tilde{t}^{(L)}$ transforms irreducibly as a
tensor of spin $L$. In the present problem we have the four vectors,
$q^\mu$, $p_3^\mu$, $\epsilon'^\mu$ and $\epsilon^\mu$ ,which are
independent on $\theta_\pi^*$. We can build the independent Lorentz
scalars:
\[
q^2,\,\,\, q\cdot \epsilon,\,\,\, q\cdot \epsilon' \,\,\,\,
\hbox{and}\,\,\,\, \epsilon\cdot\epsilon'.
\]
The available Lorentz vectors would be
\[
q^\mu,\,\,\, p_3^\mu,\,\,\, \epsilon'^\mu,\,\,\, \epsilon^\mu.
\]
With these we go first to build $L=0$ quantities. This can only be
obtained through $\tilde{t}^{(0)}$ and taking into account that
the polarization vectors $\epsilon'$ and $\epsilon$ must always be
contracted at the end,
\begin{equation}
S=\tilde{t}^{(0)}\cdot
\{I^{(0)}_1(s)(\epsilon'\cdot\epsilon)+I^{(0)}_2(s)(\epsilon'\cdot
q)(\epsilon\cdot q)\},
\end{equation}
which can be expressed through $S=\epsilon'^\mu\epsilon^\nu
S_{\mu\nu}$, with $S_{\mu\nu}=I_1(s)g_{\mu\nu}+I_2(s){q_\mu
q_\nu}/{s}$.

The $d$--wave is  more complicated, since we have to use
$\tilde{t}^{\mu\nu}_{(2)}$ and the number of contractions gets
larger. The available four-vectors,  $q^\mu, p_3^\mu, \epsilon'^\mu, \epsilon^\mu$,
are then contracted with $\tilde{t}^{\mu\nu}_{(2)}$ in all the different
possible ways:
\begin{eqnarray}
&&D=\tilde{t}^{\mu\nu}_{(2)}\, \cdot
\,  \left\{I^{(0)}_3(s)(\epsilon'\cdot\epsilon)q^\mu q^\nu
+ I^{(0)}_4(s)(\epsilon'\cdot\epsilon)p_3^\mu p_3^\nu
+ I^{(0)}_5(s) (\epsilon'\cdot\epsilon)(p_3^\mu q^\nu+ q^\mu p_3^\nu)
\right.
\nonumber\\
&&+I^{(0)}_6(s)(\epsilon'\cdot q)(\epsilon \cdot q)q^\mu q^\nu
+I^{(0)}_7(s)(\epsilon'\cdot q)(\epsilon \cdot q)p_3^\mu p_3^\nu
+I^{(0)}_8(s)(\epsilon'\cdot q)(\epsilon \cdot q)(p_3^\mu q^\nu+q^\mu p_3^\nu)
\nonumber\\
&&+I^{(0)}_9(s)(\epsilon \cdot q)(\epsilon'^\mu q^\nu+q^\mu \epsilon'^\nu)
+I^{(0)}_{10}(s)(\epsilon \cdot q)(\epsilon'^\mu p_3^\nu+ p_3^\mu \epsilon'^\nu)
\nonumber\\
&&+I^{(0)}_{11}(s)((\epsilon' \cdot q)(\epsilon^\mu q^\nu+q^\mu
\epsilon^\nu)) +I^{(0)}_{12}(s)(\epsilon' \cdot q)(\epsilon^\mu
p_3^\nu+ p_3^\mu \epsilon^\nu)
\nonumber\\
&&
\left.
+I^{(0)}_{13}(s)(\epsilon'^\mu\epsilon^\nu+\epsilon^\mu\epsilon'^\nu)
\right\}  \, .
\end{eqnarray}
It is not difficult to find that there are not actually so many
independent Lorentz structure. It can be written
in a more compact way through $D=\epsilon'^\mu\epsilon^\nu
D'_{\mu\nu}$, with
\begin{eqnarray}
&&D'^{\mu\nu}=I'_3(s)\frac{1}{s^2}g^{\mu\nu}[(p_3\cdot
p)^2-\frac{1}{3}p^2(\tilde{p}_3)^2]+I'_4(s)\frac{1}{s^3}q^\mu
q^\nu[(p_3\cdot p)^2-\frac{1}{3}p^2(\tilde{p}_3^2)]\nonumber\\
&&+I'_5(s)\frac{1}{s^2}[q^\mu p^\nu(p_3\cdot
p)-\frac{1}{3}p^2q^\mu\tilde{p}_3^\nu]+I'_6(s)\frac{1}{s^2}[p^\mu
q^\nu(p_3\cdot p)-\frac{1}{3}p^2\tilde{p}_3^\mu q^\nu]\nonumber\\
&&+I'_7(s)\frac{1}{s}[p^\mu p^\nu-\frac{1}{3}p^2\tilde{g}^{\mu\nu}].
\end{eqnarray}

In order to avoid that any form-factor becomes artificially large os small,
we extract Lorentz structures that are numerically order one.
The amplitudes are then expressed in the form
\begin{eqnarray}
&&S^{\mu\nu}=I_1(s)g^{\mu\nu}+I_2(s)\frac{1}{s}q^\mu q^\nu,\\
&&D^{\mu\nu}=I_3(s)g^{\mu\nu}[\cos^2\theta_\pi^*-\frac{1}{3}]+I_4(s)\frac{1}{s}q^\mu
q^\nu[\cos^2\theta_\pi^*-\frac{1}{3}]\nonumber\\
&&+I_5(s)\frac{1}{s^{3/2}M_\Psi}[q^\mu p^\nu(p_3\cdot
p)-\frac{1}{3}p^2q^\mu\tilde{p}_3^\nu]+I_6(s)\frac{1}{s^{3/2}M_\Psi}[p^\mu
q^\nu(p_3\cdot p)-\frac{1}{3}p^2\tilde{p}_3^\mu q^\nu]\nonumber\\
&&+I_7(s)\frac{1}{s}[p^\mu p^\nu-\frac{1}{3}p^2\tilde{g}^{\mu\nu}],
\end{eqnarray}

Until this point the derivation is completely general. Now, we make  the main
assumption: we will assume that no further rescattering occurs between the $J/\Psi$
and the $\pi\pi$ system. Hence, the phase-shift of the amplitude is due to the
$\pi\pi$ final state interaction. This allows to use the Watson theorem
for the elastic scattering region (from the practical point of view,
up to the $K\overline{K}$ threshold). The decay amplitude can be then decomposed
into partial-waves ($S$, $D$...) with their phase-shifts equal to those
in $\pi\pi$ scattering (respectively, $\delta_0$, $\delta_2$...):
\begin{equation}
T=\epsilon'^\mu\epsilon^\nu [S_{\mu\nu} e^{i \delta_0}+D_{\mu\nu}
e^{i \delta_2}].
\end{equation}
The $D$--wave is supposed to be suppressed with respect to the $J=0$ component
and higher partial waves are neglected.

\subsection{Theoretical estimation to the leading  contributions}
Starting from the effective
lagrangian in Ref.~\cite{Mannel:96}, which is constructed using
chiral symmetry and heavy quark symmetry, the amplitude has the
form,
\begin{eqnarray}
\label{leadingterm} \mathcal{A}(\Psi'\rightarrow\Psi
\pi^+\pi^-)=-\frac{4}{F_0^2}\{[\frac{g}{2}(q^2-2m_\pi^2)+g_1
E_{\pi^+}E_{\pi^-}]\epsilon'\cdot\epsilon\nonumber\\
+g_2[p_{\pi^+}^\mu p_{\pi^-}^\nu+p_{\pi^-}^\mu
p_{\pi^+}^\nu]\epsilon'_\mu \epsilon_\nu\},
\end{eqnarray}
where $p_{\pi^\pm}=(E_{\pi^\pm},\vec{p}_{\pi^\pm})$ in the $\Psi'$
rest frame. We can calculate these leading terms' contribution to
our form factors $I_i(s)$, {\begin{eqnarray}
&&I_1(s)=-\frac{4}{F_0^2}[\frac{g}{2}(s-2m_\pi^2)
+\frac{s g_1\gamma^2}{4}(1-\frac{\rho(s)^2}{3}\beta^2)+g_2\frac{s \rho(s)^2}{6}],\\
&&I_2(s)=-\frac{2s}{F_0^2}g_2(1-\frac{\rho(s)^2}{3}),\\
&&I_3(s)=\frac{1}{F_0^2}g_1 \rho(s)^2 s\beta^2\gamma^2,\\
&&I_7(s)=\frac{2s}{F_0^2}g_2\ ,
\end{eqnarray}}
with the rest being vanishing at leading order. This calculation
suggests  that the form factors $I_{4}(s)$, $I_{5}(s)$, $I_{6}(s)$
are small quantities and the theoretical prediction can be checked
by future experiments. Recall that  the fit to the data can also in
principle determine $I_i(s)$.

\subsection{Expressions for angular distribution}
For three body decays
\cite{PDG:06},
\begin{equation}
d \Gamma =\frac{1}{(2\pi)^5}\frac{1}{16M^2}|\mathcal{M}|^2
|\vec{p}_{\pi\pi}^{\,\,*}||\vec{p}_3| ds\, d\Omega_{\pi\pi}^*
d\Omega_{J/\Psi},
\end{equation}
where $(\vec{p}_{\pi\pi}^{\,\,*},\Omega_{\pi\pi}^*)$ is the momentum of
$\pi^+$ in the dipion rest frame, and $\Omega_{J/\Psi}$ is the
angle of $J/\Psi$ in the $\Psi'$ rest frame. Hence, we can express
in the form:
\begin{eqnarray}
&&\frac{d\Gamma}{ds d\cos\theta_\pi^* d\cos\theta_3
d\phi_\pi^*}\propto\sum_{k\,l\,m}G_{klm}(s)\cos^k\phi_\pi^*
\cos^l\theta_3 \cos^m
\theta_\pi^*\nonumber\\
&&+ \cos\phi_\pi^*\sin\theta_3 \cos\theta_3 \sin\theta_\pi^*
\cos\theta_\pi^* (\tilde{G}_0(s)+ \tilde{G}_2(s)
\cos^2\theta_\pi^*),
\label{eq.partialwidth}
\end{eqnarray}
where $k,l=0,2$, and $m=0,2,4$. $G_{klm}(s)$ are functions of
$I_i(s)$ and $\cos(\delta_0-\delta_2)$. If we can determine
$G_{klm}(s)$ experimentally, then we can determine $I_i(s)$ and
extract the phase
shift difference $\delta_0-\delta_2$. 

Alternatively, we can find the information in the partial
distribution. What we find in the partial distributions is of the
form:
\begin{eqnarray}
&&\frac{d^2 \Gamma}{d s d \cos\theta_\pi^*}=A_0(s)+
A_2(s)\cos^2 \theta_\pi^*+ A_4(s)\cos^4 \theta_\pi^*,
\label{eq.A}\\
&&\frac{d^2 \Gamma}{d s d \cos\theta_3}=B_0(s)+ B_2(s)\cos^2
\theta_3,
\label{eq.B} \\
&&\frac{d^2 \Gamma}{d s d \cos\phi}=C_0(s)+ C_2(s)\cos^2 \phi,
\label{eq.C}
\end{eqnarray}
and another weighted distribution,
\begin{eqnarray}
&&W[s, \cos\theta_\pi^*]=\int^{+\pi}_{-\pi} d \phi_\pi^*
\int^1_{-1}d\cos\theta_3\frac{d\Gamma}{d s d\cos\theta_\pi^*
d\cos\theta_3 d\phi_\pi^*} \cos\phi_\pi^* \cos\theta_3\nonumber\\
&&\propto\int^{+\pi}_{-\pi} d \phi_\pi^* \int^1_{-1}d\cos\theta_3
\cos^2\phi_\pi^*\sin\theta_3 \cos^2\theta_3 \sin\theta_\pi^*
\cos\theta_\pi^* (\tilde{G}_0(s)+ \tilde{G}_2(s) \cos^2\theta_\pi^*)\nonumber\\
&&=\sin\theta_\pi^* \cos\theta_\pi^* (\tilde{G}_0(s)+ \tilde{G}_2(s)
\cos^2\theta_\pi^*)\int^{+\pi}_{-\pi} d \phi_\pi^*
\int^1_{-1}d\cos\theta_3 \cos^2\phi_\pi^*\sin\theta_3
\cos^2\theta_3\nonumber\\
&&=\frac{\pi^2}{4}\sin\theta_\pi^* \cos\theta_\pi^* (\tilde{G}_0(s)+
\tilde{G}_2(s) \cos^2\theta_\pi^*).
\label{eq.W}
\end{eqnarray}
$A_i(s), B_i(s), C_i(s)$ and $\tilde{G}_i$ are functions of $I_i(s)$
and $\cos(\delta_0(s)-\delta_2(s))$. If $A_i(s), B_i(s), C_i(s)$ and
$\tilde{G}_i$ can be fitted precisely, it will not be difficult to
determine the value of $I_i$ and $\delta_0-\delta_2$ at definite
energy.

$A_i(s), B_i(s), C_i(s)$ can also be written as combinations of
$G_{klm}$,
\begin{eqnarray}
&&A_0=\frac{2\pi}{3} (6 {G_{000}}+2 {G_{020}}+3
{G_{200}}+{G_{220}}) ,  \label{eq.coefA0}\\
&&A_2=\frac{2\pi}{3} (6 {G_{002}}+2 {G_{022}}+3
{G_{202}}+{G_{222}}) ,\\
&&A_4=\frac{2\pi}{3} (6 {G_{004}}+2 {G_{024}}+3
{G_{204}}+{G_{224}}) ,\\
&&B_0=\frac{2\pi}{15} (30 {G_{000}}+10 {G_{002}}+6 {G_{004}}+15
{G_{200}}+5 {G_{202}}+3 {G_{204}}) ,\\
&&B_2=\frac{2\pi}{15} (30 {G_{020}}+10 {G_{022}}+6 {G_{024}}+15
{G_{220}}+5 {G_{222}}+3 {G_{224}}) ,\\
&&C_0=\frac{4}{45} (45 {G_{000}}+15 {G_{002}}+9 {G_{004}}+15
{G_{020}}+5 {G_{022}}+3 {G_{024}}),\\
&&C_2=\frac{4}{45} (45 {G_{200}}+15 {G_{202}}+9 {G_{204}}+15
{G_{220}}+5 {G_{222}}+3 {G_{224}}).
\label{eq.coefC2}
\end{eqnarray}
The dependence of $G_{klm}$ on form-factors $I_i(s)$ and
$\cos(\delta_0^0-\delta_0^2)$ are very complicated. The
expressions are listed in  Appendix B.

\section{Efficiency corrections}
The partial and weighted distributions in
Eqs.~(\ref{eq.A})--(\ref{eq.W}) were based on a theoretical
integration of the partial decay rate over different angular
variables. However, the experimental situation is slightly different
from this. In general, the detector is not able to cover the whole
solid angle and, moreover, the detection efficiency is not the same
in all directions but it is a rather complicate function $w(\Omega)$
~\footnote{Although a priori we will assume
$w(\Omega)=w(\theta_\pi^*,\phi_\pi^*,\theta_3)$, notice that for
asymmetric detectors the efficiency could also depend on the azimuth
angle $\phi_3$}. The partial decay rate detected in the experimental
analysis is not that in Eq.~(\ref{eq.partialwidth}) but the
efficiency corrected one,
\begin{eqnarray}
&&
\left. \frac{d\Gamma}{ds d\cos\theta_\pi^* d\cos\theta_3
d\phi_\pi^*}\right|_{\rm det.}\,\,\, =\,\,\,
w(\Omega)\,\,\, \times \,\,\,
\frac{d\Gamma}{ds d\cos\theta_\pi^* d\cos\theta_3
d\phi_\pi^*}\, .
\label{eq.correctedwidth}
\end{eqnarray}

The calculation of the corrected functions corresponding to
the  distributions in Eqs.~(\ref{eq.A})--(\ref{eq.W}) is more tedious
but it does not introduce any important complication.
In order to ease the understanding of the procedure, we present a
detailed calculation for
$d^2\Gamma/ds d\cos{\theta_\pi^*}$.  We integrate the detected partial rate in
Eq.~(\ref{eq.correctedwidth}) over $\theta_3$ and $\phi_\pi^*$ and we integrate
separately every monomial $\cos^k\phi_\pi^* \cos^l \theta_3 \cos^m \theta_\pi^*$:
\begin{eqnarray}
\left.\Frac{d^2\Gamma}{ds d\cos\theta_\pi^*}\right|_{\rm det.}
&=& \, \sum_{k,l,m} G_{klm}(s)
\Int d\cos\theta_3 d\phi_\pi^* \, \,
w(\Omega)
\cos^k\phi_\pi^* \cos^l \theta_3 \cos^m \theta_\pi^*
\nonumber \\
&&  + \, \widetilde{G}_0(s)
\Int d\cos\theta_3 d\phi_\pi^*
\, \, w(\Omega)
\cos\phi_\pi^* \sin\theta_3 \cos\theta_3 \sin\theta_\pi^* \cos\theta_\pi^*
\nonumber \\
&&  + \, \widetilde{G}_2(s)
\Int d\cos\theta_3 d\phi_\pi^*
\, \, w(\Omega)
\cos\phi_\pi^* \sin\theta_3 \cos\theta_3 \sin\theta_\pi^* \cos^3\theta_\pi^* \, .
\nonumber \\
\end{eqnarray}
Since the integral is on $\theta_3$ and $\phi_\pi^*$,
it is possible  to reexpress it in the form
\begin{eqnarray}
\left. \Frac{d^2\Gamma}{ds d\cos\theta_\pi^*}\right|_{\rm det.}
&=&  \sum_{m=0,2,4}   A_m  cos^m \theta_\pi^*
 + \sin\theta_\pi^* \cos\theta_\pi^*  \left[ \widetilde{A}_0
 +  \widetilde{A}_2 \cos^2\theta_\pi^* \right] ,
\end{eqnarray}
where we have defined a new set of coefficients $A_i,\, \widetilde{A}_j$ given by
\begin{eqnarray}
A_m &=& \, \sum_{k,l=0,2} G_{klm}(s)
\Int d\cos\theta_3 d\phi_\pi^* \, \,
w(\Omega)
\cos^k\phi_\pi^* \cos^l \theta_3 \, ,
\\
\widetilde{A}_0 &=&   \widetilde{G}_0(s)
\Int d\cos\theta_3 d\phi_\pi^*
\, \, w(\Omega)
\cos\phi_\pi^* \sin\theta_3 \cos\theta_3 \, ,
\\
\widetilde{A}_2 &=&  \widetilde{G}_2(s)
\Int d\cos\theta_3 d\phi_\pi^*
\, \, w(\Omega)
\cos\phi_\pi^* \sin\theta_3 \cos\theta_3 \, .
\end{eqnarray}
It the case of perfect efficiency, $w(\Omega)=1$, one finds
$\widetilde{A}_0=\widetilde{A}_2=0$ and the different $A_i$ become
those provided in Eqs.~(\ref{eq.coefA0})--(\ref{eq.coefC2}). The
dependence on $s$ is implicitly assumed. Furthermore, if the
efficiency depends on $\theta_\pi^*$ then the coefficients
$A_i,\,\widetilde{A}_j$ are also functions of this angle. In this
case, when analyzing the experimental data one should compute these
integrals for every $\theta_\pi^*$. The simplest procedure to
evaluate these integrals is through the Monte Carlo  method, where
we have for instance
\begin{equation}
\Int d\cos\theta_3 d\phi_\pi^* \, \,
w(\Omega)
\cos^k\phi_\pi^* \cos^l \theta_3
\,\,\, \simeq \,\,\, \Frac{1}{N_{\rm tot}} \, \sum_{a=1}^{N_{\rm MC}}
\cos^k\phi^*_{\pi,\, a}  \cos^l \theta_{3,\, a}\, .
\end{equation}
In the case of a $\theta_\pi^*$--dependent efficiency, this integral also depends
on this angle and it must be repeated for every point in the fit analysis.

Through a similar procedure, one also recovers the detected distributions
corresponding to those in Eqs.~(\ref{eq.B})--(\ref{eq.W}):
\begin{eqnarray}
\left. \Frac{d^2\Gamma}{ds d\cos\theta_3} \right|_{\rm det.}
&=&
\, \sum_{l=0,2}   B_l \, cos^l \theta_3
\, + \, \widetilde{B}\, \sin\theta_3 \cos\theta_3 \, ,
\\
\left. \Frac{d^2\Gamma}{ds d\cos\phi_\pi^*} \right|_{\rm det.}
&=&
\, \sum_{k=0,2}   C_k \, cos^k \phi_\pi^*
\, + \, \widetilde{C}\, \cos\phi_\pi^* \, ,
\\
\left. W[s,\theta_\pi^*] \right|_{\rm det.}
&=&  \sum_{m=0,2,4}   W_k  \cos^m \theta_\pi^*
 +  \sin\theta_\pi^* \cos\theta_\pi^*
 \left[ \widetilde{W}_0 + \widetilde{W}_2  \cos^2\theta_\pi^* \right] ,
\end{eqnarray}
with the coefficients
\begin{eqnarray}
B_l &=& \, \sum_{k=0,2}\sum_{m=0,2,4} G_{klm}(s)
\Int d\cos\theta_\pi^* d\phi_\pi^* \, \,
w(\Omega)
\cos^k\phi_\pi^* \cos^m \theta_\pi^* \, ,
\\
\widetilde{B} &=&
\Int d\cos\theta_\pi^* d\phi_\pi^*
\, \, w(\Omega)
\cos\phi_\pi^* \sin\theta_\pi^* \cos\theta_\pi^*  \,
[ \widetilde{G}_0(s) \, +\,
 \widetilde{G}_2(s)   \cos^2\theta_\pi^* ] \, ,
\\
C_k &=& \, \sum_{l=0,2}\sum_{m=0,2,4} G_{klm}(s)
\Int d\cos\theta_\pi^* d\cos\theta_3 \, \,
w(\Omega)
\cos^l\theta_3 \cos^m \theta_\pi^* \, ,
\\
\widetilde{C} &=&
\Int d\cos\theta_\pi^* d\cos\theta_3
\, \, w(\Omega)
\sin\theta_3 \cos\theta_3 \sin\theta_\pi^* \cos\theta_\pi^*  \,
\nonumber \\
&&\qquad \qquad \times \,\,
[ \widetilde{G}_0(s) \, +\,
 \widetilde{G}_2(s)   \cos^2\theta_\pi^* ] \, ,
\\
W_m &=& \, \sum_{k,l=0,2} G_{klm}(s)
\Int d\cos\theta_3 d\phi_\pi^* \, \,
w(\Omega)
\cos^{k+1}\phi_\pi^* \cos^{l+1} \theta_3 \, ,
\\
\widetilde{W}_0 &=&   \widetilde{G}_0(s)
\Int d\cos\theta_3 d\phi_\pi^*
\, \, w(\Omega)
\cos^2\phi_\pi^* \sin\theta_3 \cos^2\theta_3 \, ,
\\
\widetilde{W}_2 &=&  \widetilde{G}_2(s)
\Int d\cos\theta_3 d\phi_\pi^*
\, \, w(\Omega)
\cos\phi_\pi^* \sin\theta_3 \cos^2\theta_3 \, .
\end{eqnarray}
As it happened before, for a general efficiency $w(\Omega)$,
these coefficients are not simply functions of the energy but they also have
a residual dependence on the corresponding angle.

\section{Monte Carlo Study}
Clear signals of $\sigma$ and $\kappa$ are found in BES
data\cite{bes01,bes02,Jpsipipi2}. In this studies, measuring the
$\pi\pi/\pi K$ S-wave phase shift is tried, but because of the
limited statistics, no meaningful results are obtained. In the
$J/\Psi \to \omega \pi \pi$ channel, there are resonances in the
$\omega \pi$ spectrum, which will affect the S-wave phase in the
$\pi\pi$ spectrum. Its contribution to the $\pi\pi$ S-wave phase
shift is hard to be estimated theoretically, which is the trouble
for the measurement of the $\pi\pi$ S-wave phase shift in the
$J/\Psi \to \omega \pi \pi$ channel. However, all these troubles do
not exist in the $\Psi^{\prime} \to \pi \pi J/\Psi$ channel, for the
energy of the $\pi J/\Psi$ spectrum is too low and no resonances
exist in the $\pi J/\Psi$ spectrum. For BESII data, the channel
$\Psi^{\prime} \to \pi \pi J/\Psi$, where $J/\Psi \to \mu^+ \mu^-$,
is studied\cite{Jpsipipi2}, and a global partial wave analysis is
performed. After introducing a wide $0^{++}$ background which
strongly destructively interfere with $\sigma$  particle, the
$\pi\pi$ spectrum can be well fitted. The pole position measured in
this channel is consistent with that measured in the $J/\Psi \to
\omega \pi \pi$ channel. Though global PWA fit can obtain reasonable
results on $\sigma$ particle, $\pi\pi$ S-wave phase shift can not be
well  defined. The reason is that the statistics in BESII data is
too low to perform a reasonable fit on phase shift, which is studied
in this paper. In this  paper, we will use Monte-Carlo technique to
generate data with different D-wave components and different
statistics, and then use the method proposed in this paper to fit
the data.
\\

It is expected that BESIII will collect  huge number of
$\Psi^{\prime}$ events.  The statistics of BESIII data will be about
one thousand times of BESII  data. For example, the statistics of
$\Psi^{\prime} \to \pi \pi J/\Psi$ in a 10 MeV bin at 500 MeV in the
$\pi\pi$ spectrum in BESII data is about 1,000, if BESIII statistics
is 1,000 times larger, we will have about 1 million statistics in a
10 MeV bin. So, in our Monte Carlo simulation, about one million
Monte Carlo events are  generated, and the method proposed in this
paper is used to fit the  data to see whether reasonable phase shift
can be obtained or not.  For the Monte Carlo data, the S-wave phase
motion is known, so we can test the above method by comparing the
fitted results with the input value of Monte Carlo simulation.
\\

\begin{figure}[htbp]
\begin{flushleft}
{{\epsfig{file=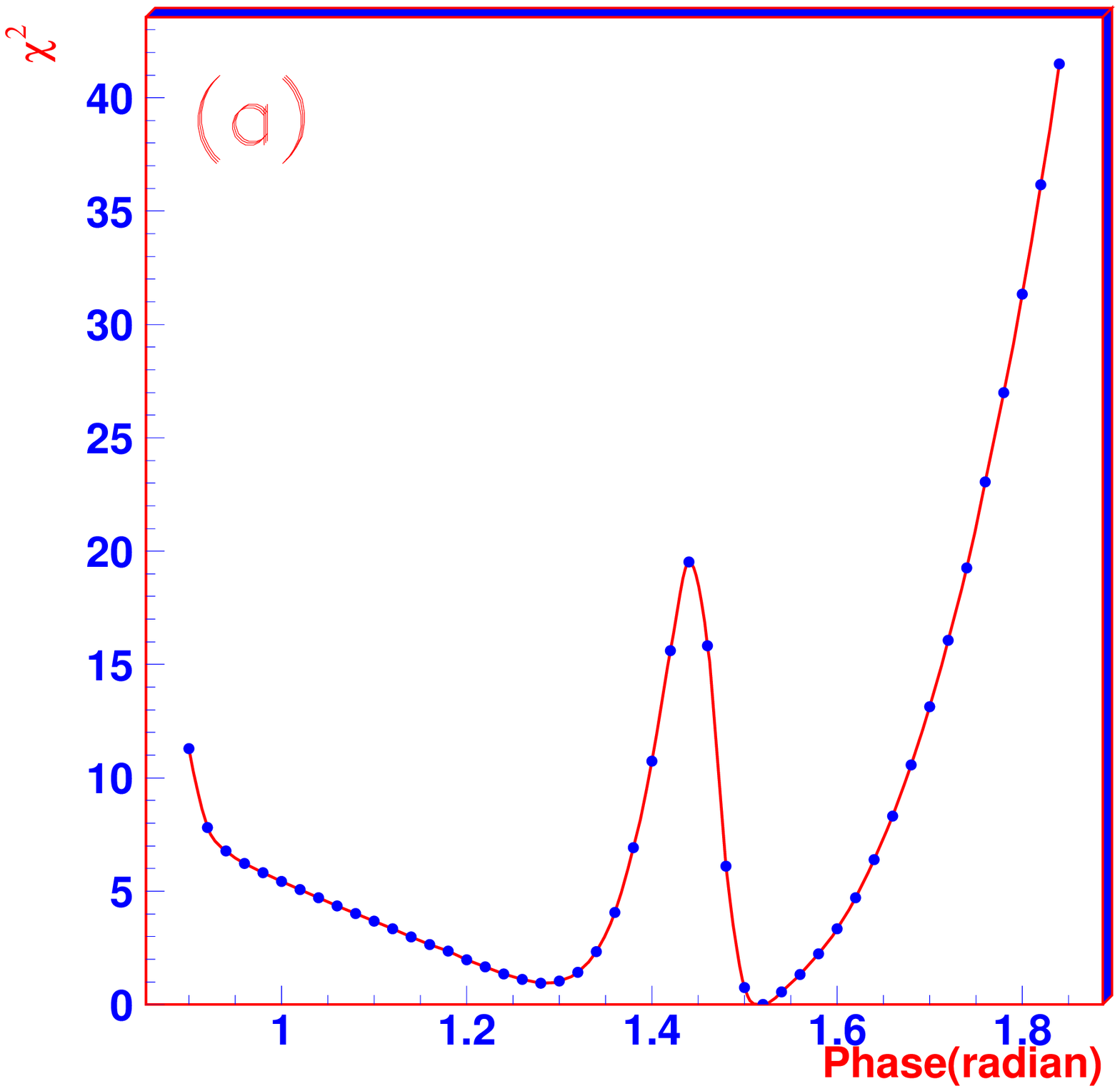,height=2.3in,width=2.8in}}}
\end{flushleft}
\vspace{-2.65in}
\begin{flushright}
{\mbox{\epsfig{file=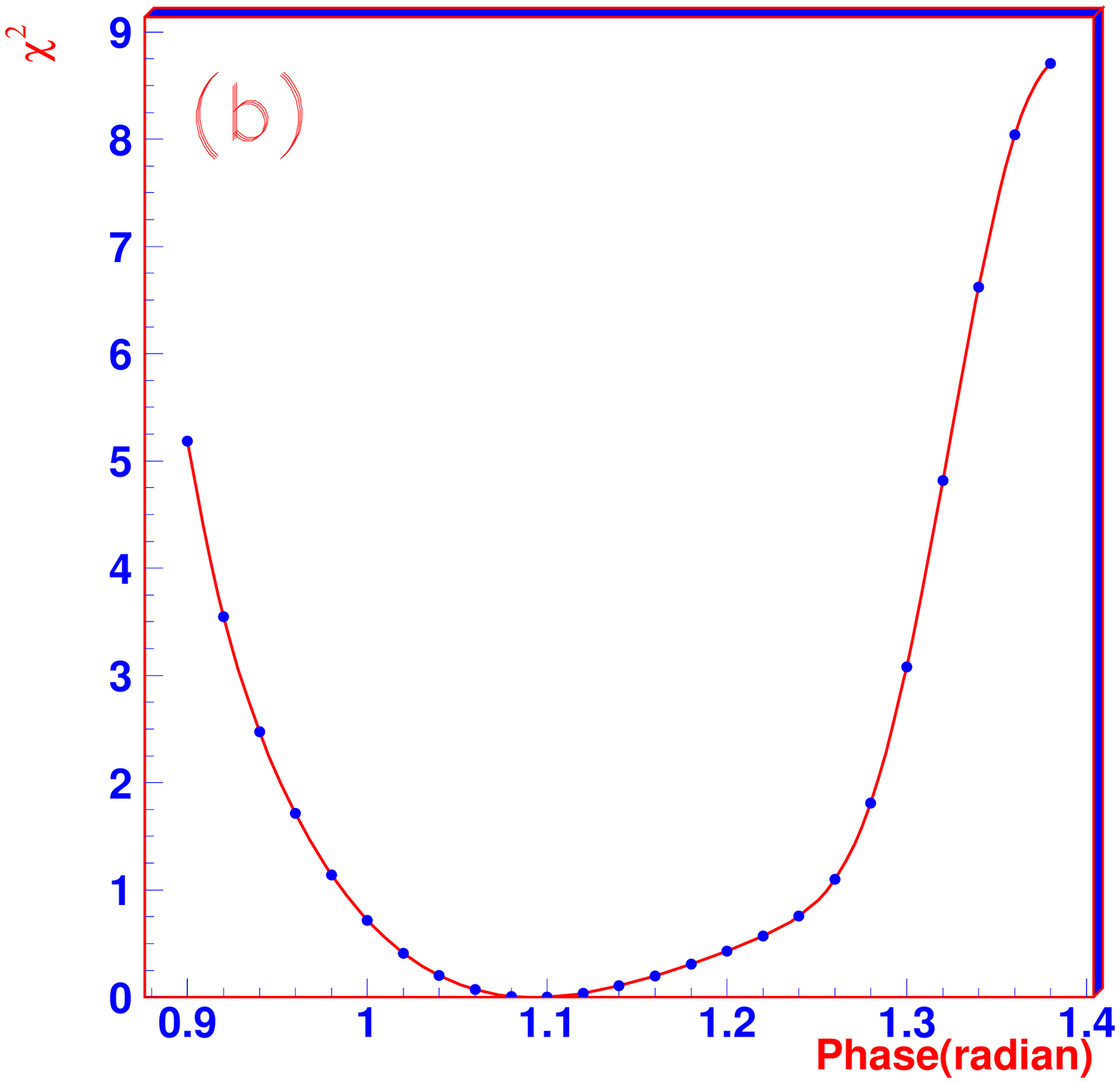,height=2.3in,width=2.8in}}}
\end{flushright}
\begin{flushleft}
{\mbox{\epsfig{file=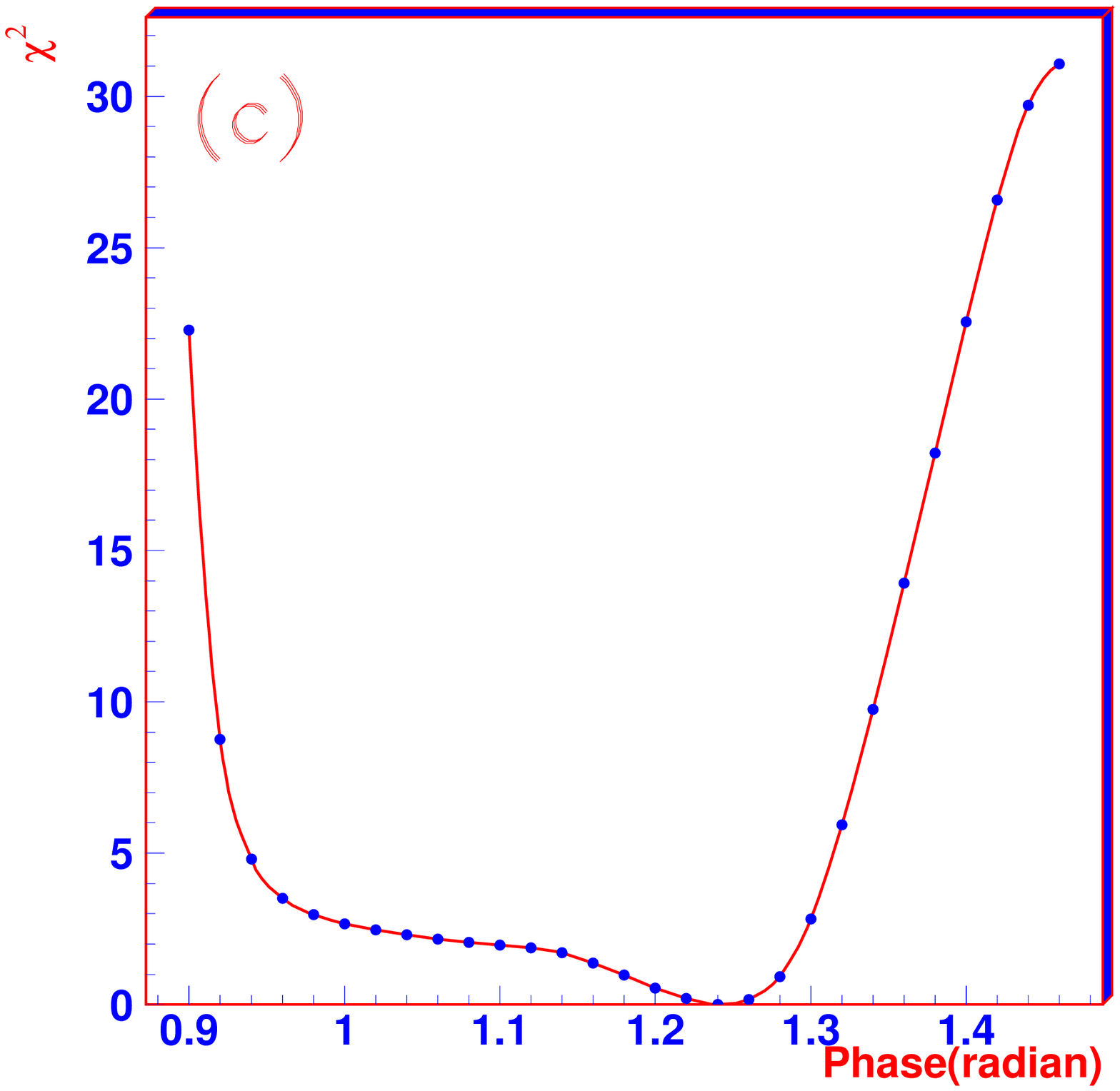,height=2.3in,width=2.8in}}}
\end{flushleft}
\vspace{-2.65in}
\begin{flushright}
{\mbox{\epsfig{file=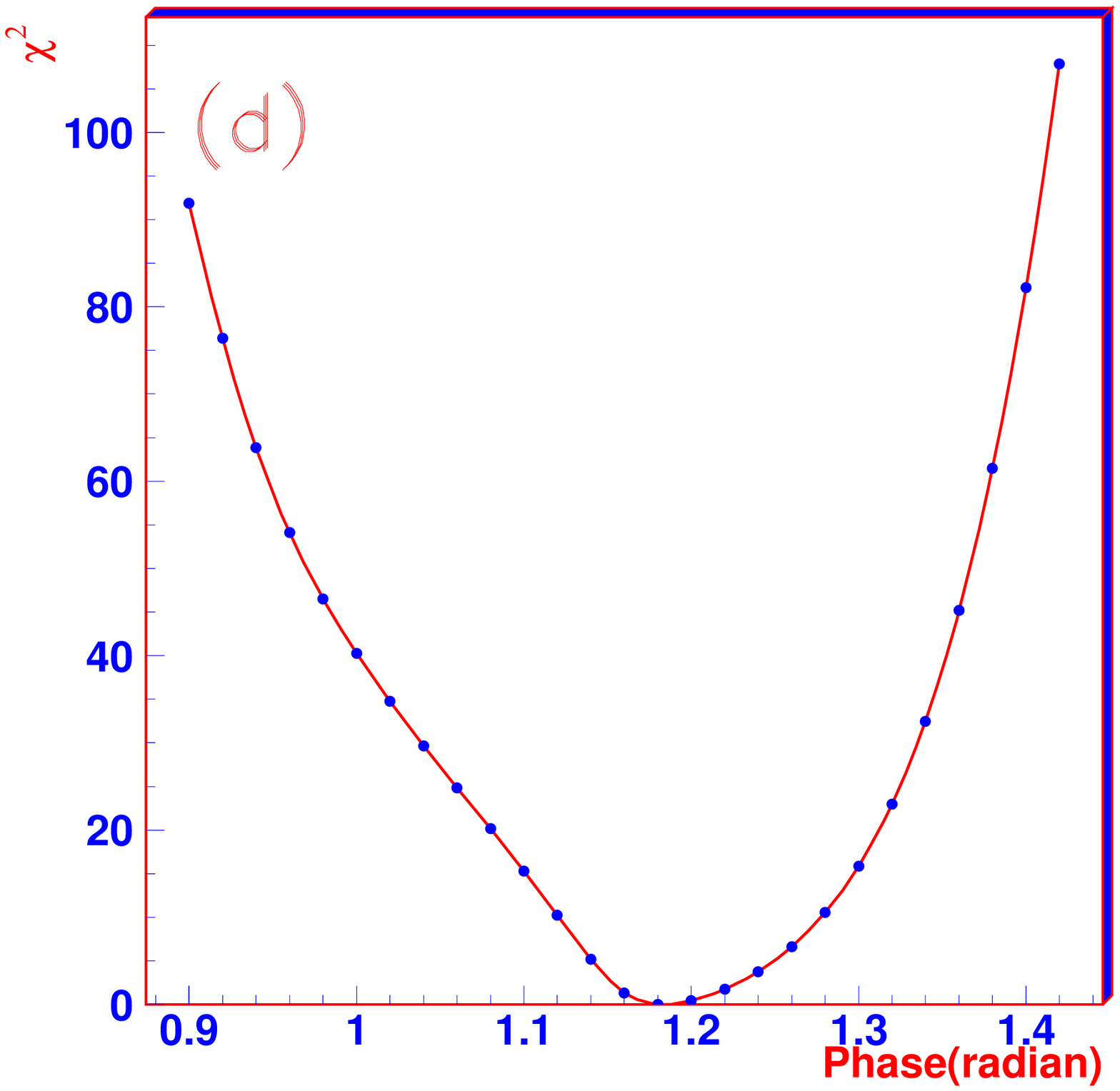,height=2.3in,width=2.8in}}}
\end{flushright}
\begin{flushleft}
{\mbox{\epsfig{file=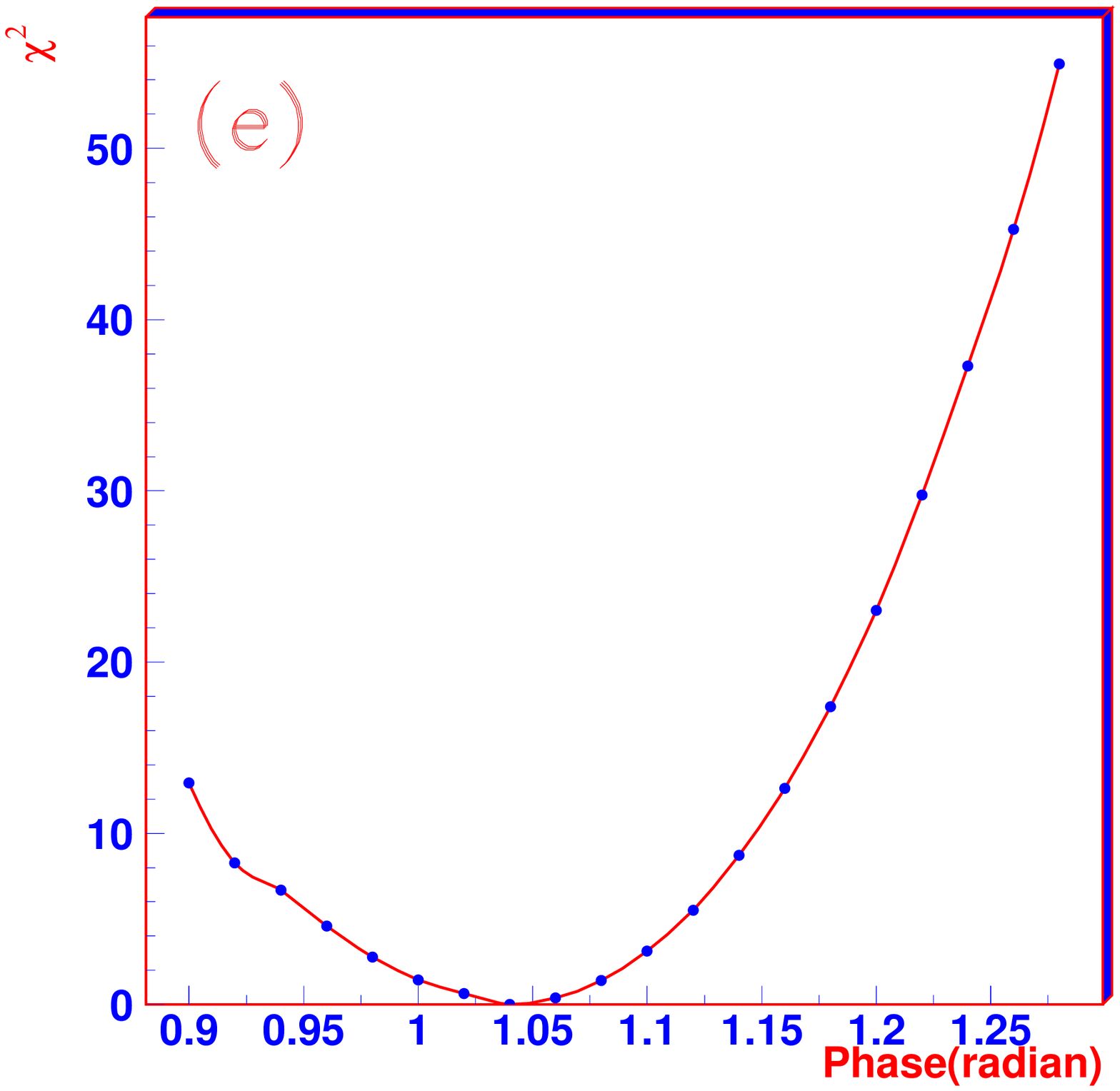,height=2.3in,width=2.8in}}}
\end{flushleft}
\caption[]{ Scan results on I,J=0,0 phase. (a), (b), (c), (d)
    and (e) are scan  results when D-component are 2\%, 4\%,
    8\%, 20\% and 45\% respectively. The value at minimum of the
    curve is the output phase. }
\label{k07}
\end{figure}

In the Monte Carlo simulation, we need first to know the amount of
D-wave component, or the percentage of D-wave component in the total
Monte Carlo data sample. According to literature \cite{Jpsipipi}, the
ratios of D-wave component to S-wave component in the $m_{\pi\pi}$
range from 340 MeV to 600 MeV are in the range  from 4.7\%  to
31.9\%, and the  ratio decreases with the increase of $m_{\pi\pi}$.
In order to simplify the problem, we generate Monte Carlo data in
the energy between 500 MeV to 510 MeV with different D-wave
component. Five independent Monte Carlo data samples are generated
with D-component 2\%, 4\%, 8\%, 20\% and 45\% respectively. In each
data sample, the method proposed in this paper ia used to fit the
data, then a scan on I,J=0,0 phase is done. Scan results on the
phases of the Monte Carlo data samples are shown in figure
\ref{k07}. In these figures, we can see that there is a minimum in
the smooth scan curve, and the phase value at the minimum is the
I,J=0,0 phase of that case. In all these cases except the 45\% case,
the input phases are 1.17, which corresponding to 67$^{\circ}$. In
the 45\% case the input phase is taken as 1.03 radian.  The fit
results are listed in table I. It can be seen that the fit results
are quite close to Monte Carlo inputs. So, the phases obtained by
the method of this paper are reasonable.

\begin{table}[htp]
\begin{center}
\doublerulesep 0pt
\renewcommand\arraystretch{1.1}
\begin{tabular}{|c|c|c|c|c|c|}
\hline
\hline
\hline

Ratio of D-wave  &  2\%  &  4\%   &  8\%   &  20\%  &  45\%  \\
\hline

$\delta^0_0-\delta^2_0$   & $1.28  \pm 0.08 $  & $1.10 \pm 0.16$
    & $1.24 \pm 0.08$  &  $1.18 \pm 0.04$
    & $1.04  \pm 0.04 $ \\

  &   $ 1.52 \pm 0.04 $ &&&& \\

\hline
\hline
\end {tabular}
\caption {Scan results on $\delta^0_0-\delta^2_0$ for different
D-wave components. Errors are only
    statistical. The input phase difference of Monte Carlo
    data are 1.03 for the 45\% case and 1.17 for the rest. }
\end{center}
\end{table}

Statistics of data is key important in the study of the  phase
difference. In the above study, the statistics of Monte Carlo data
are 600,000, 250,000, 200,000, and 200,000 for D-wave component 2\%,
4\%, 8\%, and 20\% respectively. When the ratio of D-wave component
is small, we need much higher statistics, otherwise likelihood
function is not sensitive to the change of S-wave phase, or even not
changed when S-wave phase is changed to any other value. In BESII
data, there are only about 2,000 events in a 10 MeV bin when
$m_{\pi\pi}$ is at 500 MeV. In these case, the fit is not at all
sensitive to the phase shift difference. In the Monte Carlo study,
we found similar results, that is, when D-wave component is 4\% and
statistics of the Monte Carlo data is below 10,000, the likelihood
function almost kept unchanged when we change the S-wave phase to
other value. Therefore, the reason that we can not obtain a
reasonable result on S-wave phase shift is that the statistics of
the BESII data is too few. It is expected that BESIII will collect
200 times more $\psi '$ data in the near future. And BESIII detector
has much higher selection efficiency than that of BESII detector.
So, we will have more than 400,000 events in a 10 MeV bin when
$m_{\pi\pi}$ is at 500 MeV. Our Monte Carlo study shows that, if the
ratio of D-wave component is above 2\%,  a reasonable S-wave phase
shift can be obtained based on BESIII data.
\\

\section{Conclusions}

In this paper, a method is proposed to measure $\pi\pi$ S-wave phase
shift, and this method is tested by Monte Carlo data. In the Monte
Carlo study, it is found that, even if the ratio of the D-wave
component is above 2\% and the statistics in one 10 MeV bin is above
about 200,000, a reasonable results on $\pi\pi$ S-wave phase shift
can be obtained. It is expected that BESIII will collect enough
data, so based on BESIII data, we can measure $\pi\pi$ S-wave phase
shift in the mass region
from 350 MeV to 550 MeV. \\

{\bf\large Acknowledgements}:
  This work is support in part by National
 Nature Science Foundations of China under contract number
 10575002,
  10491306 and 
 10721063, 
and by the EU-RTN Programme, Contract
No MRTN-CT-2006-035482, "Flavianet".

\begin{appendix}

\section{Kinematics}
$\epsilon'$ and $\epsilon$ denote the polarizations of $\Psi'$ and
$J/\Psi$ respectively. Here we choose $$\epsilon_0^\mu=
\frac{1}{M}(|\vec{p}_3|, 0, 0, -E_3),$$
 $$\epsilon_\pm^\mu= \frac{\sqrt{2}}{2}(0, 1, \pm i, 0),$$  and
 $$\epsilon^{'\mu}_{1}= (0, -\cos \theta_3 \cos \phi,
-\cos \theta_3 \sin \phi, -\sin \theta_3),$$
 $$\epsilon^{'\mu}_{2}= (0, -\sin \phi, \cos \phi, 0).$$
\begin{eqnarray}
&&{p_3\cdot q}=\sqrt{s}( E_3+|\vec{p}_3|\beta),\nonumber\\
&&p_3\cdot \epsilon'_{1}=-|\vec{p}_3| \sin\theta_3,\nonumber\\
&&p\cdot \epsilon'_{1}=2|\vec{p}_{\pi\pi}^*|(\cos\theta_3 \cos\phi
\sin\theta_\pi^*+\gamma \cos\theta_\pi^* \sin\theta_3),\nonumber\\
&&{q\cdot \epsilon'_{1}}=\sqrt{s}\beta\gamma \sin\theta_3,\nonumber\\
&&p_3 \cdot \epsilon'_{2}=0,\nonumber\\
&&p\cdot \epsilon'_{2}=2|\vec{p}_{\pi\pi}^*|\sin\theta_\pi^* \sin\phi,\nonumber\\
&&q\cdot\epsilon'_{2}=0,\nonumber\\
&&p\cdot\epsilon_+=-\sqrt{2}|\vec{p}_{\pi\pi}^*|\sin\theta_\pi^*,\nonumber\\
&&p\cdot\epsilon_-=-\sqrt{2}|\vec{p}_{\pi\pi}^*|\sin\theta_\pi^*,\nonumber\\
&&p\cdot\epsilon_0=2|\vec{p}_{\pi\pi}^*|\cos\theta_\pi^*\gamma(E_3+|\vec{p}_3|\beta)\frac{1}{M_\Psi},\nonumber\\
&&q\cdot\epsilon_+=0, \nonumber\\
&&q\cdot\epsilon_-=0, \nonumber\\
&&q\cdot\epsilon_0=s\gamma(|\vec{p}_3|+E_3\beta)\frac{1}{M_\Psi}\nonumber\\
&&\epsilon'_{1}\cdot\epsilon_+=\cos\theta_3(\cos\phi+i \sin\phi)/\sqrt{2}, \nonumber\\
&&\epsilon'_{1}\cdot\epsilon_-=\cos\theta_3(\cos\phi-i \sin\phi)/\sqrt{2}, \nonumber\\
&&\epsilon'_{1}\cdot\epsilon_0=-E_3 \sin\theta_3/M_\Psi,\nonumber\\
&&\epsilon'_{2}\cdot\epsilon_+=(-i\cos\phi+ \sin\phi)/\sqrt{2}, \nonumber\\
&&\epsilon'_{2}\cdot\epsilon_-=(i\cos\phi+ \sin\phi)/\sqrt{2}, \nonumber\\
&&\epsilon'_{2}\cdot\epsilon_0=0.\nonumber
\end{eqnarray}

\section{$G_{klm}$ and $\tilde{G}_{i}$}
We have used the notations
\[P_3=|\vec{p}_3|,V=\beta,M=M_{J/\Psi},\] and $\delta=\delta_0-\delta_2$
in the definitions of $G_{klm}$ and $\tilde{G}_{i}$ 
~\footnote{The Mathematica
notebook and fortran programs can be obtained from
xiaoly@pku.edu.cn    and   cillero@ifae.es. }.  
{\scriptsize
\begin{eqnarray}
&&G_{224}=\frac{1}{M^4}(2 {I_3} {I_7} \rho ^2 M^4+\rho ^4 (-{I_6}^2
({P_3}+{E_3} V)^4 \gamma ^4-2 {I_6} {I_7} M
   ({P_3}+{E_3} V)^2 ({E_3}+{P_3} V) \gamma ^3\nonumber\\
&&+{I_7}^2 M^2 (M^2-({E_3}+{P_3} V)^2 \gamma
   ^2))),\\
&&G_{204}=-G_{224},\\
&&G_{222}=\frac{1}{3M^4}(\rho ^2 (3 {I_6}^2 \rho ^2 ({P_3}+{E_3}
V)^4 \gamma ^4-{I_7}^2 M^2 \rho ^2 (4 M^2-3
   ({E_3}+{P_3} V)^2 \gamma ^2)\nonumber\\
&&+2 {I_7} M (3 {\cos \delta} {I_1} M^3-4 {I_3} M^3+3
   {I_6} \rho ^2 ({P_3}+{E_3} V)^2 ({E_3}+{P_3} V) \gamma ^3))),\\
&&G_{202}=-G_{222},\\
&&G_{220}=\frac{1}{3} {I_7} \rho ^2 ({I_7} \rho ^2-6
{\cos \delta} {I_1}+2 {I_3}),\\
&&G_{220}=-G_{200},\\
&&G_{024}=-\frac{1}{M^4}({E_3}^4 ({I_5}+{I_6})^2 \rho ^4 V^4 \gamma
^6+2 {E_3}^3 ({I_5}+{I_6}) \rho ^2 V^2 (\gamma
   ^2 ({I_7} M \rho ^2+V ({P_3} ({I_5} V^2+{I_5}+2 {I_6}) \gamma  \rho ^2+{I_4} M
   V))\nonumber\\
&&-{I_3} M) \gamma ^3+({I_7}^2 M^2 ({P_3}^2 V^2 \gamma ^2-M^2) \rho
^4+2
   {I_7} M {P_3} V \gamma  ({I_5} ({P_3}^2 V^2 \gamma ^2-M^2) \rho ^2+{P_3} \gamma
   ({I_6} {P_3} \gamma  \rho ^2+{I_4} M V)) \rho ^2\nonumber\\
&&+{P_3}^2 \gamma ^2
   ((({P_3}^2 V^4 \gamma ^2-M^2 V^2) {I_5}^2+2 {I_6} {P_3}^2 V^2 \gamma ^2
   {I_5}+{I_6}^2 {P_3}^2 \gamma ^2) \rho ^4+2 {I_4} M {P_3} V ({I_5} V^2+{I_6})
   \gamma  \rho ^2+{I_4}^2 M^2 V^2)) \gamma ^2\nonumber\\
&&+2 {E_3} (\gamma  ({I_5}^2 {P_3} V^3
   \gamma  ({P_3}^2 (V^2+1) \gamma ^2-M^2) \rho ^4+{I_5} V ({I_7} M V ({P_3}^2
   (V^2+2) \gamma ^2-M^2) \rho ^2\nonumber\\
&&+{P_3}^2 \gamma ^2 ({I_6} {P_3} (3 V^2+1) \gamma
    \rho ^2+{I_4} M (2 V^3+V))) \rho ^2+{P_3} \gamma  (({I_7}^2 V M^2+{I_6}
   {I_7} {P_3} (2 V^2+1) \gamma  M+2 {I_6}^2 {P_3}^2 V \gamma ^2) \rho ^4\nonumber\\
&&+{I_4} M V
   ({I_7} M (V^2+1)+3 {I_6} {P_3} V \gamma ) \rho ^2+{I_4}^2 M^2
   V^3))-{I_3} M {P_3} (({I_7} M V+{P_3} ({I_5} V^2+{I_6}) \gamma
   ) \rho ^2+{I_4} M V)) \gamma ^2\nonumber\\
&&-{I_3}^2 M^4+{E_3}^2 (({I_7}^2 M^2 \rho ^4+2
   {I_7} M V ({P_3} (2 {I_5} V^2+{I_5}+{I_6} (V^2+2)) \gamma  \rho ^2+{I_4}
   M V) \rho ^2\nonumber\\
&&+V^2 ((({P_3}^2 (V^4+4 V^2+1) \gamma ^2-M^2 V^2) {I_5}^2+6
   {I_6} {P_3}^2 (V^2+1) \gamma ^2 {I_5}+6 {I_6}^2 {P_3}^2 \gamma ^2) \rho ^4+2 {I_4}
   M {P_3} V (3 {I_6}\nonumber\\
&&+{I_5} (V^2+2)) \gamma  \rho ^2+{I_4}^2 M^2 V^2))
   \gamma ^4-2 {I_3} M ({I_7} M \rho ^2+V ({P_3} ({I_5} V^2+{I_5}+2 {I_6}) \gamma
    \rho ^2+{I_4} M V)) \gamma ^2+{I_3}^2
    M^2)),\\
&&G_{004}=\frac{1}{M^4}({I_6}^2 {P_3}^4 \rho ^4 \gamma ^6+{I_5}^2
{P_3}^4 \rho ^4 V^4 \gamma ^6+2 {I_5} {I_6}
   {P_3}^4 \rho ^4 V^2 \gamma ^6+2 {I_5} {I_7} M {P_3}^3 \rho ^4 V^3 \gamma ^5+2 {I_4} {I_5} M
   {P_3}^3 \rho ^2 V^3 \gamma ^5\nonumber\\
&&+2 {I_6} {I_7} M {P_3}^3 \rho ^4 V \gamma ^5+2 {I_4} {I_6} M
   {P_3}^3 \rho ^2 V \gamma ^5-{I_6}^2 {P_3}^4 \rho ^4 \gamma ^4-{I_5}^2 M^2 {P_3}^2 \rho ^4 V^2 \gamma
   ^4+{I_7}^2 M^2 {P_3}^2 \rho ^4 V^2 \gamma ^4\nonumber\\
&&+{I_4}^2 M^2 {P_3}^2 V^2 \gamma ^4+2 {I_4} {I_7} M^2
   {P_3}^2 \rho ^2 V^2 \gamma ^4+{E_3}^4 \rho ^4 V^4 ((\gamma ^2-1) {I_6}^2+2 {I_5} \gamma ^2
   {I_6}+{I_5}^2 \gamma ^2) \gamma ^4-2 {I_6} {I_7} M {P_3}^3 \rho ^4 V \gamma ^3\nonumber\\
&&-2 {I_5}
   {I_7} M^3 {P_3} \rho ^4 V \gamma ^3+2 {E_3}^3 \rho ^2 V^2 (2 {I_6}^2 {P_3} V \gamma  (\gamma
   ^2-1) \rho ^2-{I_3} ({I_5}+{I_6}) M+{I_5} \gamma ^2 ({I_7} M \rho ^2+V ({I_5}
   {P_3} (V^2+1) \gamma  \rho ^2\nonumber\\
&&+{I_4} M V))+{I_6} ({I_7} M (\gamma
   ^2-1) \rho ^2+V \gamma ^2 ({I_5} {P_3} (V^2+3) \gamma  \rho ^2+{I_4} M
   V))) \gamma ^3-{I_7}^2 M^4 \rho ^4 \gamma ^2-{I_7}^2 M^2 {P_3}^2 \rho ^4 V^2 \gamma ^2\nonumber\\
&&+2
   {E_3} ({I_7}^2 M^2 {P_3} V (\gamma ^2-1) \rho ^4+{I_7} M \gamma  ({I_6}
   {P_3}^2 (2 V^2+1) (\gamma ^2-1) \rho ^2+V ({I_5} V ({P_3}^2 (V^2+2)
   \gamma ^2-M^2) \rho ^2\nonumber\\
&&+{I_4} M {P_3} (V^2+1) \gamma )) \rho ^2-{I_3} M {P_3}
   (({I_7} M V+{P_3} ({I_5} V^2+{I_6}) \gamma ) \rho ^2+{I_4} M
   V)+{P_3} V \gamma ^2 (2 {I_6}^2 {P_3}^2 (\gamma ^2-1) \rho ^4\nonumber\\
&&+{I_6} {P_3} \gamma
   ({I_5} {P_3} (3 V^2+1) \gamma  \rho ^2+3 {I_4} M V) \rho ^2+V ({I_5}^2 V
   ({P_3}^2 (V^2+1) \gamma ^2-M^2) \rho ^4+{I_4} {I_5} M {P_3} (2 V^2+1)
   \gamma  \rho ^2\nonumber\\
&&+{I_4}^2 M^2 V))) \gamma ^2+{I_3}^2 M^4+{I_7}^2 M^4 \rho ^4+2
{I_3}
   {I_7} M^4 \rho ^2+{E_3}^2 ({I_3}^2 M^2-2 {I_3} \gamma ^2 ({I_7} M \rho ^2+V ({P_3}
   ({I_5} V^2+{I_5}+2 {I_6}) \gamma  \rho ^2\nonumber\\
&&+{I_4} M V)) M+\gamma ^2 ({I_7}^2
   M^2 (\gamma ^2-1) \rho ^4+2 {I_7} M V \gamma  ({I_6} {P_3} (V^2+2) (\gamma
   ^2-1) \rho ^2+\gamma  ({I_5} {P_3} (2 V^2+1) \gamma  \rho ^2+{I_4} M V)) \rho
   ^2\nonumber\\
&&+V^2 \gamma ^2 (6 {I_6}^2 {P_3}^2 (\gamma ^2-1) \rho ^4+{I_5}^2
({P_3}^2 (V^4+4
   V^2+1) \gamma ^2-M^2 V^2) \rho ^4+2 {I_4} {I_5} M {P_3} V (V^2+2) \gamma  \rho ^2\nonumber\\
&&+6
   {I_6} {P_3} \gamma  ({I_5} {P_3} (V^2+1) \gamma  \rho ^2+{I_4} M V) \rho
   ^2+{I_4}^2 M^2 V^2)))),\\
&&G_{022}=\frac{1}{3 M^4}(2 {I_5}^2 {P_3}^4 \rho ^4 V^4 \gamma ^6+2
{I_5} {I_6} {P_3}^4 \rho ^4 V^4 \gamma ^6+2 {E_3}^4
   ({I_5}+{I_6})^2 \rho ^4 V^4 \gamma ^6+2 {I_6}^2 {P_3}^4 \rho ^4 V^2 \gamma ^6+2 {I_5} {I_6}
   {P_3}^4 \rho ^4 V^2 \gamma ^6\nonumber\\
&&+4 {I_5} {I_7} M {P_3}^3 \rho ^4 V^3 \gamma ^5+2 {I_6} {I_7} M
   {P_3}^3 \rho ^4 V^3 \gamma ^5+4 {I_4} {I_5} M {P_3}^3 \rho ^2 V^3 \gamma ^5+2 {I_4} {I_6} M
   {P_3}^3 \rho ^2 V^3 \gamma ^5+2 {I_6} {I_7} M {P_3}^3 \rho ^4 V \gamma ^5\nonumber\\
&&+2 {I_4} {I_6} M
   {P_3}^3 \rho ^2 V \gamma ^5+2 {I_6}^2 {P_3}^4 \rho ^4 \gamma ^4+2 {I_5} {I_6} {P_3}^4 \rho ^4 V^2
   \gamma ^4-3 {I_5}^2 M^2 {P_3}^2 \rho ^4 V^2 \gamma ^4+2 {I_7}^2 M^2 {P_3}^2 \rho ^4 V^2 \gamma ^4\nonumber\\
&&+2
   {I_4}^2 M^2 {P_3}^2 V^2 \gamma ^4+4 {I_4} {I_7} M^2 {P_3}^2 \rho ^2 V^2 \gamma ^4+2 {I_6} {I_7}
   M {P_3}^3 \rho ^4 V \gamma ^3-6 {I_5} {I_7} M^3 {P_3} \rho ^4 V \gamma ^3+2 {I_4} {I_6} M
   {P_3}^3 \rho ^2 V \gamma ^3\nonumber\\
&&+2 {E_3}^3 ({I_5}+{I_6}) \rho ^2 V^2 ({I_7} M ((V^2+1)
   \gamma ^2+1) \rho ^2-2 {I_3} M+V \gamma  ({I_6} {P_3} ((V^2+3) \gamma ^2+1)
   \rho ^2+2 \gamma  ({I_5} {P_3} (V^2+1) \gamma  \rho ^2\nonumber\\
&&+{I_4} M V))) \gamma
   ^3-3 {I_7}^2 M^4 \rho ^4 \gamma ^2+2 {E_3} (\gamma  (({I_7}^2 {P_3} V
   ((V^2+1) \gamma ^2+1) M^2+{I_7} \gamma  ({I_6} (\gamma ^2 V^4+(5 \gamma
   ^2+1) V^2+2) {P_3}^2\nonumber\\
&&+{I_5} V^2 ({P_3}^2 (3 (V^2+1) \gamma ^2+1)-3
   M^2)) M+{P_3} V \gamma ^2 ({I_6}^2 ((3 V^2+1) \gamma ^2+3)
   {P_3}^2+{I_5} {I_6} (2 \gamma ^2 V^4+(5 \gamma ^2+2) V^2+\gamma ^2\nonumber\\
&&+1)
   {P_3}^2+{I_5}^2 V^2 (2 {P_3}^2 (V^2+1) \gamma ^2-3 M^2))) \rho ^4+{I_4} M
   {P_3} V ({I_7} M ((3 V^2+1) \gamma ^2+1)+2 {P_3} V \gamma  ({I_5} (2
   V^2+1) \gamma ^2\nonumber\\
&&+{I_6} ((V^2+2) \gamma ^2+1))) \rho ^2+2 {I_4}^2 M^2
   {P_3} V^3 \gamma ^2)-{I_3} M {P_3} ({I_6} {P_3} ((V^2+1) \gamma ^2+1)
   \rho ^2+2 V \gamma  (({I_7} M+{I_5} {P_3} V \gamma ) \rho ^2\nonumber\\
&&+{I_4} M))) \gamma -2
   {I_3}^2 M^4+2 {I_3} {I_7} M^4 \rho ^2+6 {\cos \delta} M (-{I_2} {P_3}^2 V
   (({I_7} M V+{P_3} ({I_5} V^2+{I_6}) \gamma ) \rho ^2+{I_4} M V) \gamma
   ^4\nonumber\\
&&+{E_3}^3 ({I_5}+{I_6}) \rho ^2 V^2 ({I_1}-{I_2} V^2 \gamma ^2)
\gamma ^3+{E_3}
   {P_3} ({I_1} (({I_7} M V+{P_3} ({I_5} V^2+{I_6}) \gamma ) \rho
   ^2+{I_4} M V)-{I_2} V (\gamma ^2 ({I_7} M (V^2\nonumber\\
&&+1) \rho ^2+V ({P_3} (2
   {I_5} V^2+{I_5}+3 {I_6}) \gamma  \rho ^2+2 {I_4} M V))-{I_3} M)) \gamma
   ^2+{I_1} {I_3} M^3+{E_3}^2 ({I_1} (\gamma ^2 ({I_7} M \rho ^2\nonumber\\
&&+V ({P_3}
   ({I_5} V^2+{I_5}+2 {I_6}) \gamma  \rho ^2+{I_4} M V))-{I_3} M)-{I_2}
   V^2 \gamma ^2 (\gamma ^2 ({I_7} M \rho ^2+V ({P_3} (3 {I_6}+{I_5}
   (V^2+2)) \gamma  \rho ^2+{I_4} M V))\nonumber\\
&&-{I_3} M)))+{E_3}^2 (2
   {I_3}^2 M^2-2 {I_3} ({I_7} M ((V^2+1) \gamma ^2+1) \rho ^2+V \gamma  ({I_6}
   {P_3} ((V^2+3) \gamma ^2+1) \rho ^2+2 \gamma  ({I_5} {P_3} (V^2+1) \gamma
    \rho ^2\nonumber\\
&&+{I_4} M V))) M+\gamma ^2 (2 {I_7}^2 M^2 (V^2 \gamma ^2+1) \rho
^4+2
   {I_7} M V ({P_3} \gamma  ({I_6} (5 V^2 \gamma ^2+\gamma ^2+3)+{I_5} (\gamma ^2
   V^4+4 \gamma ^2 V^2+V^2\nonumber\\
&&+\gamma ^2+1)) \rho ^2+{I_4} M V ((V^2+1) \gamma ^2+1))
   \rho ^2+V^2 \gamma  (\gamma  ((2 {P_3}^2 (V^4+4 V^2+1) \gamma ^2-3 M^2 V^2)
   {I_5}^2+2 {I_6} {P_3}^2 (\gamma ^2 V^4\nonumber\\
&&+(7 \gamma ^2+1) V^2+4 \gamma ^2+2) {I_5}+6
   {I_6}^2 {P_3}^2 ((V^2+1) \gamma ^2+1)) \rho ^4+2 {I_4} M {P_3} V (2
   {I_5} (V^2+2) \gamma ^2\nonumber\\
&&+{I_6} ((V^2+5) \gamma ^2+1)) \rho ^2+2 {I_4}^2
   M^2 V^2 \gamma )))),\\
&&G_{002}=-\frac{1}{3 M^4}(2 {I_5}^2 {P_3}^4 \rho ^4 V^4 \gamma ^6+2
{I_5} {I_6} {P_3}^4 \rho ^4 V^4 \gamma ^6+2 {I_6}^2
   {P_3}^4 \rho ^4 V^2 \gamma ^6+2 {I_5} {I_6} {P_3}^4 \rho ^4 V^2 \gamma ^6+4 {I_5} {I_7} M
   {P_3}^3 \rho ^4 V^3 \gamma ^5\nonumber\\
&&+2 {I_6} {I_7} M {P_3}^3 \rho ^4 V^3 \gamma ^5+4 {I_4} {I_5} M
   {P_3}^3 \rho ^2 V^3 \gamma ^5+2 {I_4} {I_6} M {P_3}^3 \rho ^2 V^3 \gamma ^5+2 {I_6} {I_7} M
   {P_3}^3 \rho ^4 V \gamma ^5+2 {I_4} {I_6} M {P_3}^3 \rho ^2 V \gamma ^5\nonumber\\
&&-{I_6}^2 {P_3}^4 \rho ^4
   \gamma ^4+2 {I_5} {I_6} {P_3}^4 \rho ^4 V^2 \gamma ^4-3 {I_5}^2 M^2 {P_3}^2 \rho ^4 V^2 \gamma ^4+2
   {I_7}^2 M^2 {P_3}^2 \rho ^4 V^2 \gamma ^4+2 {I_4}^2 M^2 {P_3}^2 V^2 \gamma ^4\nonumber\\
&&+4 {I_4} {I_7} M^2
   {P_3}^2 \rho ^2 V^2 \gamma ^4+{E_3}^4 \rho ^4 V^4 ((2 \gamma ^2-3) {I_6}^2+4 {I_5} \gamma
   ^2 {I_6}+2 {I_5}^2 \gamma ^2) \gamma ^4-4 {I_6} {I_7} M {P_3}^3 \rho ^4 V \gamma ^3-6 {I_5}
   {I_7} M^3 {P_3} \rho ^4 V \gamma ^3\nonumber\\
&&+2 {I_4} {I_6} M {P_3}^3 \rho ^2 V \gamma ^3+2 {E_3}^3 \rho ^2
   V^2 (2 {I_5}^2 {P_3} \rho ^2 V (V^2+1) \gamma ^3-2 {I_3} ({I_5}+{I_6}) M+{I_6}
   ({I_7} M ((V^2+1) \gamma ^2-2) \rho ^2\nonumber\\
&&+V \gamma  ({I_6} {P_3}
   ((V^2+3) \gamma ^2-5) \rho ^2+2 {I_4} M V \gamma ))+{I_5} ({I_7} M
   ((V^2+1) \gamma ^2+1) \rho ^2+V \gamma  ({I_6} {P_3} ((3 V^2+5) \gamma
   ^2+1) \rho ^2\nonumber\\
&&+2 {I_4} M V \gamma ))) \gamma ^3-3 {I_7}^2 M^4 \rho ^4 \gamma
^2-3
   {I_7}^2 M^2 {P_3}^2 \rho ^4 V^2 \gamma ^2+2 {E_3} (\gamma  (({I_7}^2 {P_3} V
   ((V^2+1) \gamma ^2-2) M^2+{I_7} \gamma  ({I_6} (\gamma ^2 V^4\nonumber\\
&&+5 (\gamma
   ^2-1) V^2-1) {P_3}^2+{I_5} V^2 ({P_3}^2 (3 (V^2+1) \gamma ^2+1)-3
   M^2)) M+{P_3} V \gamma ^2 ({I_6}^2 ((3 V^2+1) \gamma ^2-3)
   {P_3}^2\nonumber\\
&&+{I_5} {I_6} (2 \gamma ^2 V^4+(5 \gamma ^2+2) V^2+\gamma ^2+1)
   {P_3}^2+{I_5}^2 V^2 (2 {P_3}^2 (V^2+1) \gamma ^2-3 M^2))) \rho ^4+{I_4} M
   {P_3} V ({I_7} M ((3 V^2+1) \gamma ^2\nonumber\\
&&+1)+2 {P_3} V \gamma  ({I_5} (2
   V^2+1) \gamma ^2+{I_6} ((V^2+2) \gamma ^2+1))) \rho ^2+2 {I_4}^2 M^2
   {P_3} V^3 \gamma ^2)-{I_3} M {P_3} ({I_6} {P_3} ((V^2+1) \gamma ^2+1)
   \rho ^2\nonumber\\
&&+2 V \gamma  (({I_7} M+{I_5} {P_3} V \gamma ) \rho ^2+{I_4} M)))
\gamma +2
   {I_3}^2 M^4+4 {I_7}^2 M^4 \rho ^4+6 {I_3} {I_7} M^4 \rho ^2\nonumber\\
&&-6 {\cos \delta} M ({I_2}
   {P_3}^2 V (({I_7} M V+{P_3} ({I_5} V^2+{I_6}) \gamma ) \rho ^2+{I_4} M
   V) \gamma ^4+{E_3}^3 ({I_5}+{I_6}) \rho ^2 V^2 ({I_2} V^2 \gamma ^2-{I_1}) \gamma
   ^3\nonumber\\
&&+{E_3} {P_3} ({I_2} V (\gamma ^2 ({I_7} M (V^2+1) \rho ^2+V ({P_3}
   (2 {I_5} V^2+{I_5}+3 {I_6}) \gamma  \rho ^2+2 {I_4} M V))-{I_3}
   M)\nonumber\\
&&-{I_1} (({I_7} M V+{P_3} ({I_5} V^2+{I_6}) \gamma ) \rho
   ^2+{I_4} M V)) \gamma ^2+{I_1} M^3 ({I_7} \rho ^2+{I_3})+{E_3}^2 ({I_2}
   V^2 (\gamma ^2 ({I_7} M \rho ^2+V ({P_3} (3 {I_6}\nonumber\\
&&+{I_5} (V^2+2))
   \gamma  \rho ^2+{I_4} M V))-{I_3} M) \gamma ^2+{I_1} ({I_3} M-\gamma ^2
   ({I_7} M \rho ^2+V ({P_3} ({I_5} V^2+{I_5}+2 {I_6}) \gamma  \rho ^2+{I_4} M
   V)))))\nonumber\\
&&+{E_3}^2 (2 {I_3}^2 M^2-2 {I_3} ({I_7} M
   ((V^2+1) \gamma ^2+1) \rho ^2+V \gamma  ({I_6} {P_3} ((V^2+3) \gamma
   ^2+1) \rho ^2\nonumber\\
&&+2 \gamma  ({I_5} {P_3} (V^2+1) \gamma  \rho ^2+{I_4} M
   V))) M+\gamma ^2 ({I_7}^2 M^2 (2 V^2 \gamma ^2-1) \rho ^4\nonumber\\
&&+2 {I_7} M V
   ({P_3} \gamma  ({I_5} (\gamma ^2 V^4+(4 \gamma ^2+1) V^2+\gamma ^2+1)+{I_6}
   ((5 \gamma ^2-3) V^2+\gamma ^2-3)) \rho ^2\nonumber\\
&&+{I_4} M V ((V^2+1) \gamma
   ^2+1)) \rho ^2+V^2 \gamma  (\gamma  ((2 {P_3}^2 (V^4+4 V^2+1) \gamma ^2-3 M^2
   V^2) {I_5}^2\nonumber\\
&&+2 {I_6} {P_3}^2 (\gamma ^2 V^4+(7 \gamma ^2+1) V^2+4 \gamma ^2
   {I_5}+6 {I_6}^2 {P_3}^2 ((V^2+1) \gamma ^2-2)) \rho ^4\nonumber\\
&&+2 {I_4} M {P_3} V
   (2 {I_5} (V^2+2) \gamma ^2+{I_6} ((V^2+5) \gamma ^2+1)) \rho ^2+2
   {I_4}^2 M^2 V^2 \gamma ))))),\\
&&G_{020}=-\frac{1}{9 M^4}({I_5}^2 {P_3}^4 \rho ^4 V^4 \gamma
^6+{I_6}^2 {P_3}^4 \rho ^4 V^4 \gamma ^6+2 {I_5} {I_6}
   {P_3}^4 \rho ^4 V^4 \gamma ^6+{E_3}^4 ({I_5}+{I_6})^2 \rho ^4 V^4 \gamma ^6+2 {I_5} {I_7} M
   {P_3}^3 \rho ^4 V^3 \gamma ^5\nonumber\\
&&+2 {I_6} {I_7} M {P_3}^3 \rho ^4 V^3 \gamma ^5-6 {\cos\delta}
{I_2}
   {I_5} M {P_3}^3 \rho ^2 V^3 \gamma ^5+2 {I_4} {I_5} M {P_3}^3 \rho ^2 V^3 \gamma ^5-6 {\cos\delta
   } {I_2} {I_6} M {P_3}^3 \rho ^2 V^3 \gamma ^5\nonumber\\
&&+2 {I_4} {I_6} M {P_3}^3 \rho ^2 V^3 \gamma ^5+2
   {I_6}^2 {P_3}^4 \rho ^4 V^2 \gamma ^4+2 {I_5} {I_6} {P_3}^4 \rho ^4 V^2 \gamma ^4+{I_7}^2 M^2
   {P_3}^2 \rho ^4 V^2 \gamma ^4+9 {I_2}^2 M^2 {P_3}^2 V^2 \gamma ^4\nonumber\\
&&+{I_4}^2 M^2 {P_3}^2 V^2 \gamma ^4-6
   {\cos\delta} {I_2} {I_4} M^2 {P_3}^2 V^2 \gamma ^4-6 {\cos\delta} {I_2} {I_7} M^2
   {P_3}^2 \rho ^2 V^2 \gamma ^4+2 {I_4} {I_7} M^2 {P_3}^2 \rho ^2 V^2 \gamma ^4\nonumber\\
&&+2 {I_6} {I_7} M
   {P_3}^3 \rho ^4 V \gamma ^3-6 {\cos\delta} {I_2} {I_6} M {P_3}^3 \rho ^2 V \gamma ^3+2 {I_4}
   {I_6} M {P_3}^3 \rho ^2 V \gamma ^3+2 {E_3}^3 ({I_5}+{I_6}) \rho ^2 V^2 ({I_5} {P_3} \rho
   ^2 V^3 \gamma ^3\nonumber\\
&&+{I_6} {P_3} \rho ^2 V^3 \gamma ^3+{I_5} {P_3} \rho ^2 V \gamma
^3+{I_6} {P_3} \rho
   ^2 V \gamma ^3+{I_7} M \rho ^2 V^2 \gamma ^2+{I_4} M V^2 \gamma ^2+{I_6} {P_3} \rho ^2 V \gamma +{I_7}
   M \rho ^2\nonumber\\
&&-{I_3} M+3 {\cos\delta} M ({I_1}-{I_2} V^2 \gamma ^2)) \gamma
^3+{I_6}^2
   {P_3}^4 \rho ^4 \gamma ^2+2 {E_3} {P_3} ({I_5}^2 {P_3}^2 \rho ^4 V^5 \gamma ^5+{I_6}^2
   {P_3}^2 \rho ^4 V^5 \gamma ^5\nonumber\\
&&+2 {I_5} {I_6} {P_3}^2 \rho ^4 V^5 \gamma ^5+{I_5}^2 {P_3}^2 \rho
^4
   V^3 \gamma ^5+{I_6}^2 {P_3}^2 \rho ^4 V^3 \gamma ^5+2 {I_5} {I_6} {P_3}^2 \rho ^4 V^3 \gamma ^5+2
   {I_5} {I_7} M {P_3} \rho ^4 V^4 \gamma ^4\nonumber\\
&&+2 {I_6} {I_7} M {P_3} \rho ^4 V^4 \gamma ^4+2 {I_4}
   {I_5} M {P_3} \rho ^2 V^4 \gamma ^4+2 {I_4} {I_6} M {P_3} \rho ^2 V^4 \gamma ^4+{I_5} {I_7} M
   {P_3} \rho ^4 V^2 \gamma ^4+{I_6} {I_7} M {P_3} \rho ^4 V^2 \gamma ^4\nonumber\\
&&+{I_4} {I_5} M {P_3} \rho
   ^2 V^2 \gamma ^4+{I_4} {I_6} M {P_3} \rho ^2 V^2 \gamma ^4+{I_7}^2 M^2 \rho ^4 V^3 \gamma ^3+2 {I_6}^2
   {P_3}^2 \rho ^4 V^3 \gamma ^3+2 {I_5} {I_6} {P_3}^2 \rho ^4 V^3 \gamma ^3\nonumber\\
&&+9 {I_2}^2 M^2 V^3 \gamma
   ^3+{I_4}^2 M^2 V^3 \gamma ^3+2 {I_4} {I_7} M^2 \rho ^2 V^3 \gamma ^3+{I_6}^2 {P_3}^2 \rho ^4 V \gamma
   ^3+{I_5} {I_6} {P_3}^2 \rho ^4 V \gamma ^3+{I_5} {I_7} M {P_3} \rho ^4 V^2 \gamma ^2\nonumber\\
&&+3 {I_6}
   {I_7} M {P_3} \rho ^4 V^2 \gamma ^2+2 {I_4} {I_6} M {P_3} \rho ^2 V^2 \gamma ^2+{I_7}^2 M^2 \rho ^4
   V \gamma +{I_6}^2 {P_3}^2 \rho ^4 V \gamma -9 {I_1} {I_2} M^2 V \gamma \nonumber\\
&&+{I_4} {I_7} M^2 \rho ^2 V
   \gamma +{I_6} {I_7} M {P_3} \rho ^4-{I_3} M ({I_6} {P_3} (V^2 \gamma ^2+1) \rho
   ^2+V \gamma  (({I_7} M+{I_5} {P_3} V \gamma ) \rho ^2+{I_4} M))\nonumber\\
&&+3 {\cos\delta} M
   ({I_1} ({I_6} {P_3} (V^2 \gamma ^2+1) \rho ^2+V \gamma  (({I_7} M+{I_5}
   {P_3} V \gamma ) \rho ^2+{I_4} M))-{I_2} V \gamma  ({I_7} M (2 V^2 \gamma ^2+1)
   \rho ^2\nonumber\\
&&-{I_3} M+V \gamma  ({I_6} {P_3} ((2 V^2+1) \gamma ^2+2) \rho
^2+\gamma
   ({I_5} {P_3} (2 V^2+1) \gamma  \rho ^2+2 {I_4} M V))))) \gamma -9
   {I_1}^2 M^4\nonumber\\
&&-{I_3}^2 M^4-{I_7}^2 M^4 \rho ^4+2 {I_3} {I_7} M^4 \rho ^2+6
{\cos\delta} {I_1}
   M^4 ({I_3}-{I_7} \rho ^2)+{E_3}^2 ({I_5}^2 {P_3}^2 \rho ^4 V^6 \gamma ^6+{I_6}^2
   {P_3}^2 \rho ^4 V^6 \gamma ^6\nonumber\\
&&+2 {I_5} {I_6} {P_3}^2 \rho ^4 V^6 \gamma ^6+4 {I_5}^2 {P_3}^2
\rho ^4
   V^4 \gamma ^6+4 {I_6}^2 {P_3}^2 \rho ^4 V^4 \gamma ^6+8 {I_5} {I_6} {P_3}^2 \rho ^4 V^4 \gamma
   ^6+{I_5}^2 {P_3}^2 \rho ^4 V^2 \gamma ^6\nonumber\\
&&+{I_6}^2 {P_3}^2 \rho ^4 V^2 \gamma ^6+2 {I_5} {I_6}
   {P_3}^2 \rho ^4 V^2 \gamma ^6+2 {I_5} {I_7} M {P_3} \rho ^4 V^5 \gamma ^5+2 {I_6} {I_7} M {P_3}
   \rho ^4 V^5 \gamma ^5\nonumber\\
&&-6 {\cos\delta} {I_2} {I_5} M {P_3} \rho ^2 V^5 \gamma ^5+2 {I_4}
{I_5} M
   {P_3} \rho ^2 V^5 \gamma ^5-6 {\cos\delta} {I_2} {I_6} M {P_3} \rho ^2 V^5 \gamma ^5+2 {I_4}
   {I_6} M {P_3} \rho ^2 V^5 \gamma ^5\nonumber\\
&&+4 {I_5} {I_7} M {P_3} \rho ^4 V^3 \gamma ^5+4 {I_6} {I_7} M
   {P_3} \rho ^4 V^3 \gamma ^5-12 {\cos\delta} {I_2} {I_5} M {P_3} \rho ^2 V^3 \gamma ^5+4 {I_4}
   {I_5} M {P_3} \rho ^2 V^3 \gamma ^5\nonumber\\
&&-12 {\cos\delta} {I_2} {I_6} M {P_3} \rho ^2 V^3 \gamma ^5+4
   {I_4} {I_6} M {P_3} \rho ^2 V^3 \gamma ^5+{I_7}^2 M^2 \rho ^4 V^4 \gamma ^4+2 {I_6}^2 {P_3}^2 \rho
   ^4 V^4 \gamma ^4\nonumber\\
&&+2 {I_5} {I_6} {P_3}^2 \rho ^4 V^4 \gamma ^4+9 {I_2}^2 M^2 V^4
\gamma ^4+{I_4}^2 M^2
   V^4 \gamma ^4-6 {\cos\delta} {I_2} {I_4} M^2 V^4 \gamma ^4-6 {\cos\delta} {I_2} {I_7} M^2
   \rho ^2 V^4 \gamma ^4\nonumber\\
&&+2 {I_4} {I_7} M^2 \rho ^2 V^4 \gamma ^4+4 {I_6}^2 {P_3}^2 \rho ^4
V^2 \gamma ^4+4
   {I_5} {I_6} {P_3}^2 \rho ^4 V^2 \gamma ^4+2 {I_5} {I_7} M {P_3} \rho ^4 V^3 \gamma ^3+4 {I_6}
   {I_7} M {P_3} \rho ^4 V^3 \gamma ^3\nonumber\\
&&-6 {\cos\delta} {I_2} {I_6} M {P_3} \rho ^2 V^3 \gamma ^3+2
   {I_4} {I_6} M {P_3} \rho ^2 V^3 \gamma ^3+2 {I_5} {I_7} M {P_3} \rho ^4 V \gamma ^3+2 {I_6}
   {I_7} M {P_3} \rho ^4 V \gamma ^3\nonumber\\
&&+2 {I_7}^2 M^2 \rho ^4 V^2 \gamma ^2+{I_6}^2 {P_3}^2 \rho ^4 V^2
   \gamma ^2-6 {\cos\delta} {I_2} {I_7} M^2 \rho ^2 V^2 \gamma ^2+2 {I_4} {I_7} M^2 \rho ^2 V^2 \gamma
   ^2+2 {I_6} {I_7} M {P_3} \rho ^4 V \gamma \nonumber\\
&&+{I_7}^2 M^2 \rho ^4+9 {I_1}^2 M^2+{I_3}^2 M^2-2
   {I_3} M ({I_7} M (V^2 \gamma ^2+1) \rho ^2+V \gamma  ({I_6} {P_3}
   ((V^2+1) \gamma ^2+1) \rho ^2\nonumber\\
&&+\gamma  ({I_5} {P_3} (V^2+1) \gamma  \rho ^2-3
   {\cos\delta} {I_2} M V+{I_4} M V)))+6 {I_1} M ({\cos\delta}
   ({I_7} M (V^2 \gamma ^2+1) \rho ^2-{I_3} M\nonumber\\
&&+V \gamma  ({I_6} {P_3}
   ((V^2+1) \gamma ^2+1) \rho ^2+\gamma  ({I_5} {P_3} (V^2+1) \gamma  \rho
   ^2+{I_4} M V)))-3 {I_2} M V^2 \gamma ^2))),\\
&&G_{000}=\frac{1}{9 M^4 s}({I_5}^2 {P_3}^4 \rho ^4 s V^4 \gamma
^6+{I_6}^2 {P_3}^4 \rho ^4 s V^4 \gamma ^6+2 {I_5} {I_6}
   {P_3}^4 \rho ^4 s V^4 \gamma ^6+{E_3}^4 ({I_5}+{I_6})^2 \rho ^4 s V^4 \gamma ^6\nonumber\\
&&-6 {\cos\delta}
   {I_2} {I_5} M {P_3}^3 \rho ^2 s V^3 \gamma ^5+2 {I_4} {I_5} M {P_3}^3 \rho ^2 s V^3 \gamma ^5-6
   {\cos\delta} {I_2} {I_6} M {P_3}^3 \rho ^2 s V^3 \gamma ^5+2 {I_4} {I_6} M {P_3}^3 \rho ^2 s
   V^3 \gamma ^5\nonumber\\
&&+2 {I_6}^2 {P_3}^4 \rho ^4 s V^2 \gamma ^4+2 {I_5} {I_6} {P_3}^4
\rho ^4 s V^2 \gamma ^4+9
   {I_2}^2 M^2 {P_3}^2 s V^2 \gamma ^4+{I_4}^2 M^2 {P_3}^2 s V^2 \gamma ^4\nonumber\\
&&-6 {\cos\delta} {I_2}
   {I_4} M^2 {P_3}^2 s V^2 \gamma ^4-6 {\cos\delta} {I_2} {I_6} M {P_3}^3 \rho ^2 s V \gamma ^3+2
   {I_4} {I_6} M {P_3}^3 \rho ^2 s V \gamma ^3\nonumber\\
&&+2 {E_3}^3 ({I_5}+{I_6}) \rho ^2 V^2 (-{I_3} M
   s+3 {\cos\delta} M ({I_1}-{I_2} V^2 \gamma ^2) s+V \gamma  ({P_3} ({I_5} s
   (V^2+1) \gamma ^2+{I_6} s ((V^2+1) \gamma ^2+1)) \rho ^2\nonumber\\
&&+{I_4} M s V \gamma
   )) \gamma ^3+{I_6}^2 {P_3}^4 \rho ^4 s \gamma ^2+2 {E_3} {P_3} s ({P_3}^2 V \gamma
   ({I_5}^2 V^2 (V^2+1) \gamma ^4+{I_5} {I_6} (2 \gamma ^2 V^4+2 (\gamma ^2+1)
   V^2+1) \gamma ^2\nonumber\\
&&+{I_6}^2 ((V^4+V^2) \gamma ^4+(2 V^2+1) \gamma ^2+1)) \rho
   ^4+M {P_3} ({I_4} V^2 ({I_5} (2 V^2+1) \gamma ^2+{I_6} (2 V^2 \gamma ^2+\gamma
   ^2+2)) \gamma ^2\nonumber\\
&&-{I_3} ({I_5} V^2 \gamma ^2+{I_6} V^2 \gamma ^2+{I_6})+3
   {\cos\delta} ({I_1} ({I_5} V^2 \gamma ^2+{I_6} V^2 \gamma ^2+{I_6})-{I_2} V^2
   \gamma ^2 ({I_5} (2 V^2+1) \gamma ^2\nonumber\\
&&+{I_6} (2 V^2 \gamma ^2+\gamma
   ^2+2)))) \rho ^2+M^2 V \gamma  (9 {I_2}^2 V^2 \gamma ^2+{I_4}^2 V^2 \gamma
   ^2-{I_3} {I_4}+{I_1} (3 {\cos\delta} {I_4}-9 {I_2})\nonumber\\
&&+3 {\cos\delta} {I_2}
   ({I_3}-2 {I_4} V^2 \gamma ^2))) \gamma +9 {I_1}^2 M^4 s+{I_3}^2 M^4 s-6
   {\cos\delta} {I_1} {I_3} M^4 s+{I_7}^2 M^2 \rho ^4 s (4 M^2\nonumber\\
&&+({E_3} V^2 \gamma ^2+{P_3}
   V \gamma ^2+{E_3})^2)+2 {I_7} M \rho ^2 (-{I_3} M ({E_3} s ({E_3} V^2 \gamma
   ^2+{P_3} V \gamma ^2+{E_3})-2 M^2 s)\nonumber\\
&&+({P_3}+{E_3} V) \gamma  ({E_3} V^2 \gamma
   ^2+{P_3} V \gamma ^2+{E_3}) ({I_6} s ({P_3} V^2 \gamma ^2+{E_3} V \gamma
   ^2+{P_3}) \rho ^2+s V \gamma  ({I_5} ({E_3}+{P_3} V) \gamma  \rho ^2+{I_4} M))\nonumber\\
&&-3
   {\cos\delta} M ({I_2} s V ({P_3}+{E_3} V) \gamma ^2 ({E_3} V^2 \gamma ^2+{P_3} V
   \gamma ^2+{E_3})-{I_1} ({E_3} s ({E_3} V^2 \gamma ^2+{P_3} V \gamma
   ^2+{E_3})-2 M^2 s)))\nonumber\\
&&+{E_3}^2 (9 {I_1}^2 s M^2+{I_3}^2 s M^2-2 {I_3} V
   \gamma  ({P_3} ({I_5} s (V^2+1) \gamma ^2+{I_6} s ((V^2+1) \gamma
   ^2+1)) \rho ^2\nonumber\\
&&+({I_4}-3 {\cos\delta} {I_2}) M s V \gamma ) M-6 {I_1} (3 {I_2}
   M s V^2 \gamma ^2+{\cos\delta} s ({I_3} M-V \gamma  ({I_6} {P_3} ((V^2+1)
   \gamma ^2+1) \rho ^2\nonumber\\
&&+\gamma  ({I_5} {P_3} (V^2+1) \gamma  \rho ^2+{I_4} M
   V)))) M+V^2 \gamma ^2 ({P_3}^2 s ({I_5}^2 (V^4+4 V^2+1) \gamma ^4\nonumber\\
&&+2
   {I_5} {I_6} (V^2+(V^4+4 V^2+1) \gamma ^2+2) \gamma ^2+{I_6}^2 ((V^4+4
   V^2+1) \gamma ^4+2 (V^2+2) \gamma ^2+1)) \rho ^4\nonumber\\
&&-2 (3 {\cos\delta} {I_2}-{I_4})
   M {P_3} V \gamma  ({I_5} s (V^2+2) \gamma ^2+{I_6} s ((V^2+2) \gamma
   ^2+1)) \rho ^2\nonumber\\
&&+(9 {I_2}^2-6 {\cos\delta} {I_4} {I_2}+{I_4}^2) M^2 s V^2
   \gamma ^2))),\\
&&\tilde{G}_0=-\frac{1}{3M^4}(2 \rho ^2 \gamma  (-2 {I_7}^2 \rho ^2
M^4-{I_3} {I_7} M^4-2 {I_5} {I_7} {P_3} \rho ^2 V
   \gamma  M^3-{I_3} {I_5} {P_3} V \gamma  M^3+{I_4} {I_7} {P_3}^2 V^2 \gamma ^2 M^2\nonumber\\
&&+{I_7}^2
   {P_3}^2 \rho ^2 V^2 \gamma ^2 M^2+{I_5} {I_7} {P_3}^3 \rho ^2 V^3 \gamma ^3 M+{I_6} {I_7}
   {P_3}^3 \rho ^2 V^3 \gamma ^3 M+{I_4} {I_6} {P_3}^3 V \gamma ^3 M+{I_6} {I_7} {P_3}^3 \rho ^2 V
   \gamma ^3 M\nonumber\\
&&+{I_6} {I_7} {P_3}^3 \rho ^2 V \gamma  M+3 {\cos \delta} ({I_6}
V^2 \gamma
   ({I_1}-{I_2} V^2 \gamma ^2) {E_3}^3+({I_1} ({I_7} M+2 {I_6} {P_3} V \gamma
   )-{I_2} V^2 \gamma ^2 ({I_7} M+3 {I_6} {P_3} V \gamma )) {E_3}^2\nonumber\\
&&+({I_1} ({I_6}
   \gamma  {P_3}^2+{I_7} M V {P_3}+{I_5} M^2 V^2 \gamma )-{I_2} {P_3} V \gamma ^2
   ({I_7} M (V^2+1)+3 {I_6} {P_3} V \gamma )) {E_3}\nonumber\\
&&-{I_2} {P_3}^2 V \gamma
   ^2 ({I_7} M V+{I_6} {P_3} \gamma )+{I_1} M^2 ({I_7} M+{I_5} {P_3} V \gamma ))
   M+{E_3}^4 {I_6} ({I_5}+{I_6}) \rho ^2 V^4 \gamma ^4+{I_6}^2 {P_3}^4 \rho ^2 V^2 \gamma ^4\nonumber\\
&&+{I_5}
   {I_6} {P_3}^4 \rho ^2 V^2 \gamma ^4+{I_6}^2 {P_3}^4 \rho ^2 \gamma ^2+{E_3}^3 V^2 \gamma
   ({I_5} {I_7} M \gamma ^2 \rho ^2+{I_6}^2 {P_3} V \gamma  ((V^2+3) \gamma ^2+1)
   \rho ^2-{I_3} {I_6} M\nonumber\\
&&+{I_6} ({I_7} M ((V^2+1) \gamma ^2+1) \rho ^2+V \gamma ^2
   ({I_5} {P_3} (V^2+3) \gamma  \rho ^2+{I_4} M V)))+{E_3}^2
   ({I_7}^2 M^2 (V^2 \gamma ^2+1) \rho ^2\nonumber\\
&&-{I_3} M ({I_7} M+2 {I_6} {P_3} V \gamma )+3
   {I_6} {P_3} V^2 \gamma ^2 ({I_6} {P_3} ((V^2+1) \gamma ^2+1) \rho ^2+\gamma
   ({I_5} {P_3} (V^2+1) \gamma  \rho ^2+{I_4} M V))\nonumber\\
&&+{I_7} M V \gamma
   ({I_6} {P_3} ((5 V^2+1) \gamma ^2+3) \rho ^2+\gamma  ({I_5} {P_3} (2
   V^2+1) \gamma  \rho ^2+{I_4} M V)))+{E_3} ({I_7}^2 M^2 {P_3} V
   ((V^2+1) \gamma ^2+1) \rho ^2\nonumber\\
&&-{I_3} M ({I_6} \gamma  {P_3}^2+{I_7} M V
   {P_3}+{I_5} M^2 V^2 \gamma )+{I_6} {P_3}^2 V \gamma ^2 ({I_6} {P_3} ((3
   V^2+1) \gamma ^2+3) \rho ^2\nonumber\\
&&+\gamma  ({I_5} {P_3} (3 V^2+1) \gamma  \rho ^2+3 {I_4}
   M V))+{I_7} M \gamma  ({I_6} {P_3}^2 (\gamma ^2 V^4+(5 \gamma ^2+1)
   V^2+2) \rho ^2\nonumber\\
&&+V ({I_5} V ({P_3}^2 (V^2+2) \gamma ^2-2 M^2) \rho ^2+{I_4} M
   {P_3} (V^2+1) \gamma ))))),\\
&&\tilde{G}_2=\frac{1}{M^4}(2 \rho ^2 \gamma  (-{I_7}^2 \rho ^2
M^4-{I_5} {I_7} {P_3} \rho ^2 V \gamma  M^3-{I_3}
   ({I_7} M+{I_5} {P_3} V \gamma ) M^3+{I_4} {I_7} {P_3}^2 V^2 \gamma ^2 M^2\nonumber\\
&&+{I_7}^2 {P_3}^2
   \rho ^2 V^2 \gamma ^2 M^2+{I_5} {I_7} {P_3}^3 \rho ^2 V^3 \gamma ^3 M+{I_4} {I_6} {P_3}^3 V \gamma
   ^3 M+2 {I_6} {I_7} {P_3}^3 \rho ^2 V \gamma ^3 M+{E_3}^4 {I_6} ({I_5}+{I_6}) \rho ^2 V^4 \gamma
   ^4\nonumber\\
&&+{I_6}^2 {P_3}^4 \rho ^2 \gamma ^4+{I_5} {I_6} {P_3}^4 \rho ^2 V^2
\gamma ^4+{E_3}^2 (\gamma
   ^2 ({I_7}^2 M^2 \rho ^2+3 {I_6} {P_3} V^2 \gamma  ({P_3} ({I_5} V^2+{I_5}+2
   {I_6}) \gamma  \rho ^2+{I_4} M V)\nonumber\\
&&+{I_7} M V ({P_3} (2 {I_5} V^2+{I_5}+2
   {I_6} (V^2+2)) \gamma  \rho ^2+{I_4} M V))-{I_3} M ({I_7} M+2 {I_6}
   {P_3} V \gamma ))\nonumber\\
&&+{E_3}^3 V^2 \gamma  (\gamma ^2 ({I_5} ({I_7} M+{I_6} {P_3} V
   (V^2+3) \gamma ) \rho ^2+{I_6} (2 {I_7} M \rho ^2+V (4 {I_6} {P_3} \gamma  \rho
   ^2+{I_4} M V)))-{I_3} {I_6} M)\nonumber\\
&&+{E_3} (\gamma  ({I_5} V
   ({I_6} {P_3}^3 (3 V^2+1) \gamma ^3+{I_7} M V ({P_3}^2 (V^2+2) \gamma
   ^2-M^2)) \rho ^2+{P_3} \gamma  (2 ({I_7}^2 V M^2+{I_6} {I_7} {P_3} (2
   V^2+1) \gamma  M\nonumber\\
&&+2 {I_6}^2 {P_3}^2 V \gamma ^2) \rho ^2+{I_4} M V ({I_7} M
   (V^2+1)+3 {I_6} {P_3} V \gamma )))-{I_3} M ({I_6} \gamma
   {P_3}^2+{I_7} M V {P_3}+{I_5} M^2 V^2 \gamma
   )))).
\end{eqnarray}
}

\end{appendix}

\end{document}